\documentclass[12pt]{article}
\usepackage{amsfonts}
\usepackage{graphicx}
\usepackage{epstopdf}
\usepackage{epsfig}
\usepackage{amssymb}
\usepackage{setspace}
\usepackage{caption}
\usepackage{amsfonts}
\usepackage{color}
\usepackage{amsmath}
\usepackage{float}
\usepackage{comment}
\newcommand {\be}{\begin{equation}}
\newcommand {\ee}{\end{equation}}
 \newcommand {\bea}{\begin{array}}
 
 \newcommand {\eea}{\end{array}}

 \newcommand {\sch}{Schwarzschild~}
   \newcommand {\nn}{\nonumber}
\textwidth=6.5in \hoffset=-.75in \voffset=-1in \numberwithin{equation}{section}
\numberwithin{figure}{section}

\begin{document}

\begin{titlepage}
	%\bigskip \begin{flushright}
	%\\
	%\end{flushright}
	%\maketitle
	\vspace{1cm}
	\begin{center}
		{\Large \bf {Meissner effect and holographic dual for the Melvin-Kerr-Newman-Taub-NUT spacetimes}}\\
	\end{center}
	\vspace{1cm}
	\begin{center}
		\renewcommand{\thefootnote}{\fnsymbol{footnote}}
		Masoud Ghezelbash$^1${\footnote{amg142@campus.usask.ca}}, Haryanto M. Siahaan$^2${\footnote{haryanto.siahaan@unpar.ac.id}}\\ \vspace*{1cm}$^1$Department of Physics and Engineering Physics,\\
		University of Saskatchewan, Saskatoon, Saskatchewan S7N 5E2, Canada\vspace*{0.25cm}
	\\
		$^2$Jurusan Fisika, Universitas Katolik Parahyangan,\\Jalan Ciumbuleuit 94, Bandung, Jawa Barat, 40141, Indonesia
		\renewcommand{\thefootnote}{\arabic{footnote}}
	\end{center}
	\vspace*{1cm}
	\begin{abstract}
We investigate correspondence between the Melvin-Kerr-Newman-Taub-NUT spacetime, and the conformal field theory. The near horizon geometry of extremal Melvin-Kerr-Newman-Taub-NUT spacetime possesses the  $SL\left(2,{\mathbb R}\right)\times U\left(1\right)$ isometry, which allows one to employ the asymptotic symmetry group method, in establishing the holographic correspondence for the spacetime. We also use the alternative way of the stretched horizon method, to compute the extremal entropy of the black hole. We find perfect agreement between the results of asymptotic symmetry group and the stretched horizon methods, to establish the holography for the extremal Melvin-Kerr-Newman-Taub-NUT black hole. The black hole Meissner effect for the Melvin-Kerr-Taub-NUT black hole is also discussed.
	\end{abstract}
\end{titlepage}\onecolumn
%%\pacs{04.60.-m,04.62.+v,04.70.-s,04.70.dy,11.25.-w}
\bigskip

\section{Introduction}
\label{sec:intro}

The recent reports on the black hole interaction with the external magnetic field \cite{EventHorizonTelescope:2021srq,EventHorizonTelescope:2021bee} have motivated further studies on the topic \cite{Gimeno-Soler:2021ifv}-\cite{Gupta:2021vww}. There are at least two ways for modeling the back holes immersed in an external magnetic field. In the first way, which was proposed by Wald \cite{Wald:1974np}, the magnetic field outside the event horizon is weak, and the solution for the Maxwell field is constructed by using the Killing vector of the black holes. In the second way, we embed the black holes in the Melvin magnetic universe \cite{ernst-wild}. In the latter approach,  a Harrison type transformation maps a known axial symmetric and stationary black hole solution in the Einstein-Maxwell theory, into a new magnetized black hole solution. Different aspects of the black holes in the magnetic spacetimes have been extensively studied \cite{Aliev:1989wx}-\cite{Hiscock:1980zf}. 

Taub-NUT spacetime is an exact solution to the vacuum Einstein equations. This solution can be considered as an extension of \sch solution, where it contains the NUT parameter $l$, in addition to the mass parameter $m$. Nevertheless, the presence of NUT parameter leads to the so called closed timelike curve (CTC) and the conic singularity \cite{Griffiths:2009dfa}. These obscure features lead to the conclusion that Taub-NUT spacetime has no meaningful real astrophysics application, even though some revivals in connecting NUT parameter with astronomical data can be found recently in literature \cite{Zhang:2021pvx,Ghasemi-Nodehi:2021ipd}. Moreover, the solution have interesting thermodynamical aspects, where one can find the conserved quantities from the available approaches for the Taub-NUT geometry or its generalization in Einstein-Maxwell theory \cite{Pradhan:2013hqa}-\cite{Aliev:2008wv}. 

One of the most general exact solution in Einstein-Maxwell theory is known as the Kerr-Newman-Taub-NUT (KNTN) spacetime, parameterized by the mass $m$, electric charge $q$, rotation $a$, and NUT charge $l$. Indeed, due to the existence of semi-infinite line singularity on its rotational axis, and its regularity at origin, the KNTN spacetime cannot be considered to contain a well defined traditional black hole. However, studies on aspects of the black hole with the NUT parameter are available in literature \cite{Feng:2020tyc,Sakti:2017pmt,HossainAli:2013ydh,Ali:2013doa,Kerner:2006vu}, anticipating black holes with NUT parameter can really exist in the universe.

On the other hand, the Melvin magnetic universe is also an exact solution to the Einstein-Maxwell theory, describing a spacetime filled by an homogeneous magnetic field \cite{Griffiths:2009dfa}. The external magnetic field strength is parameterized by $b$, where the absence of $b$ leads to the Minkowski spacetime. Black holes can be embedded in the Melvin universe, such as magnetized \sch black hole \cite{Ernst} and further generalization, such as magnetized Kerr-Newman  \cite{ernst-wild}. The external magnetic field deforms the horizon surface \cite{Wild:1980zz}, but leaves the total area and even the Hawking temperature of the black hole unchanged, compared to the non-magnetized counterpart. The magnetized black hole solutions can be obtained by employing the Harrison transformation to a known seed solution. Such process is sometime called as the Ernst magnetization. For example, the magnetized Kerr-Newman spacetime can be obtained by Ernst magnetizing the Kerr-Newman solution. Recently, magnetization of solutions in Kerr-Newman-Taub-NUT family have been investigated in  \cite{Siahaan:2021uqo}-\cite{Siahaan:2021ypk}. Similar to the spirit of studying spacetime with the NUT charge, exploring the magnetized solutions is motivated by the theoretical considerations such as thermodynamics and the conserved quantities, rather than practical aspects.
 
The Kerr/CFT correspondence \cite{Guica:2008mu}-\cite{Compere:2012jk}, states that extremal black hole entropy can be computed by using the two dimensional conformal field theory (CFT) through the Cardy entropy. The Kerr/CFT correspondence has been generalized to some magnetized black holes \cite{Siahaan:2021uqo, Siahaan:2021ags},
\cite{Astorino:2015naa}-\cite{Siahaan:2015xia}, where the entropy of the extremal magnetized black hole, can be computed by the dual CFT, using the asymptotic symmetry group (ASG)  \cite{Guica:2008mu}. In fact, there also exist an alternative method to ASG, known as the stretched horizon method (SHM) \cite{Carlip:1998wz,Chen:2011wm} whose results for black hole entropy are in agreement to that of ASG method. 

In this paper, we investigate the holography for the Melvin-Kerr-Newman-Taub-NUT (MKNTN) spacetime \cite{Ghezelbash:2021lcf}. In particular, we employ both the ASG and SHM to find the dual CFT quantities. In ASG method, we first show the $SL\left(2,{\mathbb R}\right) \times U\left(1\right)$ isometry \cite{Hartman:2008pb,Compere:2012jk} for  the near horizon extremal MKNTN spacetime. It turns out that the form of extremal near horizon metric and the gauge field, belong to the class of the cases discussed in \cite{Compere:2012jk}. We obtain the central charge from the general formula in \cite{Compere:2012jk}. In SHM, we use the general approach presented in \cite{Carlip:1998wz,Chen:2011wm} to find the CFT dual quantities for the MKNTN spacetime.

In addition to the Kerr/CFT correspondence studies for the Melvin-Kerr-Newman-Taub-NUT spacetime, we also investigate the black hole Meissner effect that is reported in \cite{Bicak:2015lxa}. This effect is understood as the expulsion of axial component of the near horizon vector solution in the static limit of near horizon geometry at extremality. The authors of \cite{Bicak:2015lxa} show that a particular magnitude of external magnetic field that yields the static limit of near horizon of extremal magnetized Kerr black hole leads to the vanishing of axial component of the corresponding vector solution. Therefore, it is straightforward to ask whether such effect still occurs if the spacetime contains NUT parameter.

The organization of this paper, where we consider the natural units $c={\hbar} = k_B = G_4 = 1$, is as follows. In section \ref{sec.microentropy}, we consider the ASG method to establish the holography for the extremal MKNTN spacetime. In section \ref{sec:Carlip}, we consider the SHM and calculate the dual CFT quantities. We find both ASG and SHM lead to the same consistent results for the dual CFT quantities. The black hole Meissner effect for Melvin-Kerr-Taub-NUT black hole is discussed in section \ref{sec.Meissner}. We wrap up the article by conclusions.

\section{Microscopic entropy for extremal black holes}\label{sec.microentropy}

In this section, we consider the ASG method to establish the holographic dual CFT, for the MKNTN black hole. Some aspects of MKNTN solution are given in the appendix \ref{sec:MKNTN}. We consider the following transformation 
\be 
t \to \frac{{r_0 t'}}{\lambda }~~,~~r \to r_e  + \lambda r_0 r'~~,~~\phi  \to \phi'  + \Omega _J^{ext} \frac{{r_0 }}{\lambda }t',\label{trans}
\ee 
to find the near horizon geometry for the MKNTN black hole.
In (\ref{trans}), $r_e = m$ is the horizon radius for extremal black hole, $\Omega_J = \omega$ is the angular velocity, and $\Omega _J^{ext}$ is the corresponding quantity,  at extremality, where $m^2+l^2 = a^2 + q^2$.  Applying the near horizon transformation (\ref{trans}) to the metric (\ref{metricmag}), we obtain the near horizon line element
\be \label{nhmetric}
{\rm{d}}s^2  = \Gamma \left( x \right)\left\{  - r^2 {\rm{d}}t'^2  + \frac{{{\rm{d}}r'^2 }}{{r'^2 }} + \alpha \left( x \right) {{\rm{d}}x^2 } \right\} + \gamma \left( x \right)\left( {{\rm{d}}\phi'  + kr{\rm{d}}t'} \right)^2  \,,
\ee 
after setting the scale $r_0 = \sqrt{2 a^2 +q^2} $. In equation (\ref{nhmetric}), we have $\alpha\left(x\right) =\Delta_x^{-1}$ and
\[
\Gamma \left( x \right) = x^2 \left[{a}^{2} b^4 \left(9{q}^{4}+24{a}^{2}{q}^{2}+4{a}^{2}{l}^{2}+16{a}^{4}+4{l}^{2}{q}^{2}-4{l}^{4} \right) + 4 b^3 amq \left( 3{q}^{2}-2{l}^{2}+4{a}^{2} \right)\right.
\]
\[
\left. +b^2 \left(4{q}^{4}-8{a}^{4}-2{a}^{2}{q}^{2}\right) -4 abmq +a^2 \right] - 2lx \left[ a b^4 \left( 12{a}^{2}{l}^{2}+4{l}^{2}{q}^{2}-5{q}^{4}-8{a}^{2}{q}^{2}+4{l}^{4} \right) \right.
\]
\[
\left. +8 b^3 mq \left( {a}^{2}+{l}^{2} \right) +6a{b}^{2}{q}^{2}+4bmq-a \right] + b^4 \left( {a}^{2}+{q}^{2} \right)  \left( 16{a}^{4}+8{a}^{2}{q}^{2}+4{a}^{2}{l}^{2}+{q}^{4} -4{l}^{4}\right.
\]
\be \left. +4{l}^{2}{q}^{2} \right) 
+ 4 b^3 amq \left( {q}^{2}+4{a}^{2}+2{l}^{2} \right) + 2b^2 \left(4{a}^{4}+7{a}^{2}{q}^{2}+{q}^{4}\right) + 4abmq+{a}^{2}+{q}^{2}\,,
\ee 

\be 
\gamma \left( x \right) = \frac{\left(2 a^2 + q^2\right)^2 \Delta_x}{\Gamma\left(x\right)}\,,
\ee 
and
\[
k = -\frac{2}{r_0^2} \left\{ \left( 16{a}^{5}m+4{l}^{4}am+3m{q}^{4}a+4a{l}^{2}m{q}^{2}+12{a}^{3}m{l}^{2}+16{a}^{3}m{q}^{2} \right) {b}^{4}-am  \right.
\]
\be\label{k} 
\left. + 2q \left( {q}^{4}+2{l}^{2}{q}^{2}-4{a}^{2}{l}^{2}-4{l}^{4}+8
{a}^{4}+8{a}^{2}{q}^{2} \right) {b}^{3}+6a{b}^{2}m{q}^{2} -2q \left( 2{l}^{2}-{q}^{2} \right)  b \right\}.
\ee

The near horizon vector field $A_\mu$ associated to the near horizon geometry, can be obtained as  \cite{Bicak:2015lxa}
\[
{\bf A} = A_t dt + A_\phi  d\phi 
\]
\be\label{nearA}
= \left( {\left. {A_t } \right|_{r_e }  + \Omega _J^{{\rm{ext}}} \left. {A_\phi  } \right|_{r_e } } \right)\frac{{r_0 dt'}}{\lambda } + \left( {\left. {\frac{{\partial A_t }}{{\partial r}}} \right|_{r_e }  + \Omega _J^{{\rm{ext}}} \left. {\frac{{\partial A_\phi  }}{{\partial r}}} \right|_{r_e } } \right)r_0^2 r'dt' + \left. {A_\phi  } \right|_{r_e } d\phi '\,,
\ee 
where we expand the vector field, up to the first order in $\lambda$. Note that the constant term in the last equation can be gauged away by the gauge transformation ${\bf A} \to {\bf A}  + d\Lambda$  \cite{Compere:2012jk}, where
\be 
\Lambda = \frac{\Phi_e^{\rm ext}}{\lambda} r_0 t\,.
\ee 
In last equation, $\Phi_e = - \left. {\xi^\mu A_\mu}\right|_{r_+}$ is the electrostatic potential on horizon, and $\xi^\mu$ is the Killing vector associated to the MKNTN spacetime, i.e. $\xi ^\mu  \partial _\mu   = \partial _t  + \Omega _H \partial _\phi  $. The resulting near horizon vector solution reads
\be \label{nhA}
A_{\mu} dx^\mu = L\left(x\right) \left(k r' dt' + d\phi' \right) + C d\phi' \,,
\ee 
where $L\left(x\right)$ is given in the appendix \ref{app.Lx}, and again the constant $C$ can be gauged away. 

The near horizon geometry (\ref{nhmetric}) possesses the $SL\left(2,{\mathbb R}\right)\times U\left(1\right)$ isometry, which hints for the existence of holographic dual CFT. The $SL\left(2,{\mathbb R}\right)$ symmetry is generated by the Killing vectors
\be 
K _ -   = \partial _t \,,
\ee 
\be 
K _0  = t\partial _t  - r\partial _r \,,
\ee 
\be 
K _ +   = \left( {\frac{1}{{2r^2 }} + \frac{{t^2 }}{2}} \right)\partial _t  - tr\partial _r  - \frac{k}{r}\partial _\phi  \,,
\ee 
obeying the commutation relations $\left[ {K _0 ,K _ \pm  } \right] =  \pm K _ \pm   $ and $\left[ {K_ -  ,K_ +  } \right] = K_0 $, while the $U\left(1\right)$ symmetry is generated by $ \partial_\phi$. 

We also note that the solutions (\ref{nhmetric}) and (\ref{nhA}) are in agreement with the general forms of near horizon solutions, presented in \cite{Hartman:2008pb}. Therefore, we can use the general formula for the central charge of the holographic dual CFT. Furthermore, we need to impose the following boundary conditions for the spacetime metric
\be \label{bcg}
h_{\mu \nu }  \sim \left( {\begin{array}{*{20}c}
		{{\cal O}\left( {r^2 } \right)} & {{\cal O}\left( {1 } \right)} & {{\cal O}\left( {r^{ - 1} } \right)} & {\cal O}\left( r^{-2} \right)  \\
		{} & {{\cal O}\left(1\right)} & {{\cal O}\left( {r^{ - 1} } \right)} & {{\cal O}\left( {r^{ - 1} } \right)}  \\
		{} & {} & { {\cal O}\left( {r^{ - 1} } \right)} & {{\cal O}\left( {r^{ - 2} } \right)}  \\
		{} & {} & {} & {{\cal O}\left( r^{-3} \right)}  \\
\end{array}} \right) \,,
\ee 
whereas for the Maxwell vector field, we impose \cite{Hartman:2008pb,Compere:2012jk}
\be \label{bcA}
a_\mu  {\rm{d}}x^\mu   \sim {\cal O}\left( r \right){\rm{d}}t + {\cal O}\left( {r^{ - 1} } \right){\rm{d}}r + {\cal O}\left( 1 \right){\rm{d}}x + {\cal O} \left(r^{-2}\right) {\rm{d}}\phi \,.
\ee 
Accordingly, the most general diffeomorphisms preserving the boundary conditions (\ref{bcg}) and (\ref{bcA}), can be read as
\be 
K_\varepsilon   = \varepsilon \left( \phi  \right)\partial _\phi   - r\frac{{d\varepsilon \left( \phi  \right)}}{{d\phi }}\partial _r  + {\text{subleading terms}}\,.
\ee 
The central charge can be computed using the general formulas in \cite{Hartman:2008pb}
\be \label{central.charge.gen}
c = c_{grav}  + c_{gauge} \,,
\ee
where
\be \label{c.grav}
c_{grav}  = 3k\int\limits_{ - 1}^{ + 1} {dx\sqrt {\Gamma \left( x \right)\alpha \left( x \right)\gamma \left( x \right)} } \,,
\ee 
and
\be 
c_{gauge} = 0.
\ee 
Inserting the metric component (\ref{nhmetric}) into eq. (\ref{c.grav}), we find 
\[
c =24q \left( 8{a}^{4}-4{a}^{2}{l}^{2}+8{a}^{2}{q}^{2}-4{l}^{
	4}+2{l}^{2}{q}^{2}+{q}^{4} \right) {b}^{3} -12am-24q \left( 2{l}^{2}-{q}^{2} \right) b+72am{q}^{2}{b}^{2} 
\]
\be \label{central.charge}
 +12am \left( 16{a}^{4}
+12{a}^{2}{l}^{2}+16{a}^{2}{q}^{2}+4{l}^{4}+4{l}^{2}{q}^{2}+3
{q}^{4} \right) {b}^{4} .
\ee 

As we have mentioned before, the Hawking temperature for the magnetized black hole is the same as original black hole. The same result is repeated here where the Hawking temperature for the MKNTN black hole is exactly the same as the Hawking temperature for the KNTN black hole. We note that the Hawking temperature vanishes at extremal condition. However, the near horizon Frolov-Thorne temperature is non-zero, even in extremal condition. Using the result in (\ref{chem.pot}), the Frolov-Thorne temperature can be computed as 
\be \label{Tp}
T_\phi   = \mathop {\lim }\limits_{r_ +   \to m} \frac{{T_H }}{{\Omega _J^{{\rm{ext}}}  - \Omega _J }} =  - \left. {\frac{{{{\partial T_H } \mathord{\left/
					{\vphantom {{\partial T_H } {\partial r_ +  }}} \right.
					\kern-\nulldelimiterspace} {\partial r_ +  }}}}{{{{\partial \Omega _J } \mathord{\left/
					{\vphantom {{\partial \Omega _J } {\partial r_ +  }}} \right.
					\kern-\nulldelimiterspace} {\partial r_ +  }}}}} \right|_{r_ +   = m} \,.
\ee 
For MKNTN black hole, we find
\be\label{FTtemp}
T_\phi   = \frac{1}{2\pi k}\,,
\ee 
where the constant $k$ is given in (\ref{k}). 

To find the microscopic entropy, we use the Cardy formula, which reads
\be\label{Cardy}
S_{\rm Cardy} = \frac{\pi^2}{3} c T\,.
\ee 
Plugging the central charge $c$ (\ref{central.charge}) and the temperature (\ref{FTtemp}) into (\ref{Cardy}), we recover exactly the entropy of the extremal MKNTN black hole, which is given by 
\be 
S_{\rm ext.} = \frac{{\cal A}_{\rm ext.}}{4} = \pi \left(2 a^2 + q^2\right),\label{SBH}
\ee
where the area $\cal A$ is given in (\ref{area}). This result can be considered as a generalization of the Kerr/CFT holography proposal \cite{Guica:2008mu} to the case of MKNTN spacetime \cite{Ghezelbash:2021lcf}. 

Above, we have shown how the extremal Kerr/CFT prescription in \cite{Guica:2008mu} can apply to the near horizon of extremal MKNTN spacetime. However, one may try to Ernst transform the near horizon of extremal seed solution and use it to compute the central charge. The near horizon of extremal KNTN metric and vector solutions are the $b\to 0$ limit of (\ref{nhmetric}) and (\ref{nhA}), respectively. Accordingly, the associated Ernst potentials that belong to the near horizon of extremal KNTN solution can be written as
\be \label{Ersnt.Pot.grav.nhkntn}
{\cal E}_{\rm ne} = \frac{{l}^{3}-3a{l}^{2}x+ \left( 3{a}^{2}-{m}^{2} \right) l-{a}^{3}x-3a{m}^{2}x
	+ im\left\{ 2alx+3{a}^{2}+3{l}^{2}+{m}^{2} \right\}}{ax + l + im} 
\ee 
for the gravitational potential, and 
\be \label{Ersnt.Pot.EM.nhkntn}
\Phi_{\rm ne} = \frac{mqx+iq\left(lx-a\right)}{ax + l + im}
\ee 
for the electromagnetic one. Note that the extremal condition $m^2+l^2 = q^2 + a^2$ applies to the last two equations. 

Interestingly, the potentials (\ref{Ersnt.Pot.grav.nhkntn}) and (\ref{Ersnt.Pot.EM.nhkntn}) can be obtained in two ways. First from the metric and vector solutions describing the near horizon of extremal KNTN spacetime. Second, by taking the near horizon $r \to m$ and extremal limits in the corresponding Ernst potentials for the KNTN solution. It turns that out both ways yield the same potentials, as it can be understood that scalars do not change under coordinate transformations. Performing the Ernst magnetization to the near horizon and extremal potentials (\ref{Ersnt.Pot.grav.nhkntn}) and (\ref{Ersnt.Pot.EM.nhkntn}) above gives
\be \label{Ernst.Pot.Mag}
{\cal E'}_{\rm ne} = \Lambda_{\rm ne}^{-1} {\cal E}_{\rm ne}~~~{\rm and}~~~\Phi'_{\rm ne} = \Lambda_{\rm ne}^{-1} \left(\Phi_{\rm ne} - b {\cal E}_{\rm ne} \right)
\ee 
where
\[
\Lambda_{\rm ne} = \frac{1}{ax + l + im}\left\{ {b}^{2}{l}^{3}-3a{b}^{2}{l}^{2}x+ \left( 3{a}^{2}{b}^{2}-{b}^{2}{m}^{2}+1 \right) l-{b}^{2}{a}^{3}x-3a{b}^{2}{m}^{2}x \right.
\] 
\be \label{Lambda.mag}
\left. -2qbmx+ax + i \left(3{b}^{2}{l}^{2}m+ \left( 2a{b}^{2}mx-2bqx \right) l+3{a}^{2}{b}^{2}m+{b}^{2}{m}^{3}+2qab+m
\right) \right\}\,.
\ee 
It can be understood that the Ernst potentials in eqs. (\ref{Ernst.Pot.Mag}) correspond to the metric (\ref{nhmetric}) and vector (\ref{nhA}). Furthermore, the near horizon metric and vector solution describing the near horizon of extremal MKNTN spacetime can be extracted from these potentials. Therefore, it can be an alternative to obtain the solutions in eqs. (\ref{nhmetric}) and (\ref{nhA}) in addition to applying the conditions in eq. (\ref{trans}) to the MKNTN solution.

\section{Stretched horizon method}\label{sec:Carlip}

In this section, we review the stretched horizon formalism and apply it to the MKNTN spacetime (\ref{metricmag}). In ADM fomalism, we write the metric (\ref{metricmag}), as
\be
ds^2=-N^2dt^2+g_{rr}dr^2+g_{xx}dx^2+g_{\phi\phi}(\tilde N dt+d\phi)^2,\label{ADM}
\ee
where $N$ and $\tilde N$ are the lapse and shift functions. Moreover, we note that the metric functions $f$, $\omega$, $\gamma$ and $\rho$, in (\ref{metricmag}), where are given by (\ref{ff}), (\ref{oomega}), (\ref{ggama}),  and (\ref{rrho}), do not depend explicitly on time coordinate $t$ and also $\phi$. Hence all the metric functions in (\ref{ADM}) are independent of $t$ and $\phi$.  In the absence of the magnetic field, we have
\be
g_{rr0}=\frac{e^{2\gamma _0}}{f_0\Delta _r},\label{grr0}
\ee
where we denote the relevant metric functions with a subscript $0$. Moreover, for the KNTN metric, we have
\be
{g_{rr}}_0=\frac{\Sigma}{\Delta_r},\label{grr02}
\ee
where $\Sigma=r^2+(ax+l)^2$, and consequently one can show
\be
e^{2\gamma _0}=\Sigma f_0.
\ee

Furthermore, in the presence of magnetic field, we find the following equation
\be
g_{rr}=\frac{e^{2\gamma _0}}{f\Delta _r}=\frac{\Sigma f_0}{f\Delta_r}.\label{grr}
\ee
We notice for the extremal black hole $g_{rr}$ has a double pole at the outer horizon $r_+$, so we can write it as 
\be
g_{rr}=\frac{g_1(r,x)^2}{(r-r_+)^2},
\ee
where $g_1=\sqrt{\frac{\Sigma f_0}{f}}$. Explicitly, the expression of $g_1$ function is given in (\ref{g1}). In the extremal limit, the shift function $\tilde N$ is related to the angular velocity of horizon $\Omega_H$ \cite{Ghezelbash:2021lcf}, by $ \tilde N=-\Omega_H+g_2(r-r_+)$, where function $g_2$ is given in eq. (\ref{g2}). Moreover, we find the lapse function $N=\frac{\rho}{\sqrt{f}}$, can be written as a linear function of $r-r_+$, by $N=g_3(r,x) \times (r-r_+) $, where the full expression of $g_3$ is given in  (\ref{g3}).

Furnished by the above results, we can proceed now to calculate the central charge of the dual CFT. In fact, the Poisson bracket of Hamiltonians (as the generators of symmetry for the diffeomorphisms) becomes a Virasoro algebra with a central charge \cite{Carlip:1998wz}, where the central charge is given by
\be
c=\frac{3}{2\pi}{\cal A}g.\label{c}
\ee
In the last equation, the area $\cal A$ is given by (\ref{area}) and the factor $g$ in (\ref{c}) is equal to $g=\frac{g_1g_2}{g_3}$, where $g_1,g_2$ and $g_3$ are presented in (\ref{g1}), (\ref{g2}) and (\ref{g3}), respectively. We find the central charge (\ref{c}) turns out to be the same as (\ref{central.charge}) obtained by ASG method. Obviously, using the Cardy entropy (\ref{Cardy}) with the central charge (\ref{c}) and the temperature (\ref{Tp}) leads to the Bekenstein-Hawking entropy (\ref{SBH}) for the extremal MKNTN black hole. This result confirms the Kerr/CFT holography for MKNTN spacetime in SHM method.

\section{Meissner effect}\label{sec.Meissner}

The black hole Meissner effect is understood as the vanishing of external vacuum axially symmetric stationary magnetic fields as the black holes approaches extremality \cite{Bicak:2015lxa}. At extremality, it turns out that the condition for static near horizon of the magnetized Kerr black hole also implies the vanishing of axial component for the associated near horizon vector field. One may wonder if such effect can also occur in the magnetized Taub-NUT spacetime. The presence of NUT charge in the spacetime contributes to a particular kind of angular momentum with a specific conserved quantities associated to it. 

To study the Meissner effect in MKNTN spacetime, let us consider the near horizon of extremal black hole geometry given in (\ref{nhmetric}). However, instead of using the accompanying near horizon vector solution (\ref{nhA}), we prefer to employ the equivalent alternative one in the form of 
\be \label{An}
A_\mu  dx^\mu   = L\left( x \right)r'dt' + \left. {A_\phi  } \right|_{r_e } d\phi '\,,
\ee 
where the extremal condition $m^2+l^2 = a^2 +q^2$ applies. Explicitly, the axial component of the near horizon vector field (\ref{An}) reads
\be 
\left. {A_\phi  } \right|_{r_e }  = \frac{{\sum\limits_{j = 0}^5 {\beta _j l^j } }}{{\sum\limits_{k = 0}^5 {\tilde \beta _k l^k } }}\,,
\ee 
where
\[ 
\beta_5 = 8a{b}^{3}x\,,
\] 
\[ 
\beta_4 = 4{b}^{3} \left( {q}^{2}+{a}^{2}{x}^{2}+{a}^{2} \right) \,,
\] 
\[ 
\beta_3 = 4{b}^{2}x \left( 2ab{q}^{2}+3mq+6b{a}^{3} \right) \,,
\] 
\[ 
\beta_2 = -2{b}^{2} \left( 3amq+4{a}^{2}b{q}^{2}+2b{q}^{4}-3amq{x}^{2} +2{a}^{4}b{x}^{2}+2{a}^{4}b+2{a}^{2}b{q}^{2}{x}^{2} \right)\,,
\] 
\[ 
\beta_1 = -2qx \left( 5a{b}^{3}{q}^{3}+8q{a}^{3}{b}^{3}-3abq-m-6{a}^{2}{b}^{2}m \right)\,,
\] 
\[
\beta_0 = \left\{amq-{a}^{2} \left( 4{a}^{2}+3{q}^{2} \right) ^{2}{b}^{3}-3amq \left( 4{a}^{2}+3{q}^{2} \right) {b}^{2}+ \left( {a}^{2}{q}^{2}+4 {a}^{4}-2{q}^{4} \right) b\right\} {x}^{2}
\]
\[
- \left( {a}^{2}+{q}^{2} \right)  \left( {q}^{2}+4{a}^{2} \right) ^{2}{b}^{3}-3amq \left( {q}^{2}+4{a}^{2} \right) {b}^{2}- \left( {q}^{4}+4{a}^{4}+7{a}^{2}{q}^{2} \right) b-amq\,,
\]
\[
{\tilde\beta}_5 = -8 a{b}^{4}x\,,
\]
\[
{\tilde\beta}_4 = -4{b}^{4} \left( {q}^{2}+{a}^{2}{x}^{2}+{a}^{2} \right) \,,
\]
\[
{\tilde\beta}_3 = -8{b}^{3}x \left( ab{q}^{2}+3 b{a}^{3}+2 mq \right) \,,
\]
\[
{\tilde\beta}_2 = 4{b}^{3} \left( 2amq+2{a}^{2}b{q}^{2}+{a}^{4}b{x}^{2}-2amq{x}^{2}+b{q}^{4}+{a}^{2}b{q}^{2}{x}^{2}+{a}^{4}b \right)  \,,
\]
\[
{\tilde\beta}_1 = 2x \left(5a{b}^{4}{q}^{4} -6a{b}^{2}{q}^{2}-8{a}^{2}{b}^{3}mq +8{a}^{3}{b}^{4}{q}^{2}-4bmq+a \right) \,,
\]
\[
{\tilde\beta}_0 = \left\{{a}^{2} \left( 4{a}^{2}+3{q}^{2} \right) ^{2}{b}^{4}+4amq \left( 4{a}^{2}+3{q}^{2} \right) {b}^{3}+ \left( -2{a}^{2}{q}^{2}-8{a}^{4}+4{q}^{4} \right) {b}^{2}-4abmq+{a}^{2}
\right\} x^2 
\]
and
\[
+\left( {a}^{2}+{q}^{2} \right)  \left( {q}^{2}+4{a}^{2} \right) ^{2}{b}^{4}+4amq \left( {q}^{2}+4{a}^{2} \right) {b}^{3}+ \left( 8{a}^{4}+2{q}^{4}+14{a}^{2}{q}^{2} \right) {b}^{2}+4abmq+{a}^{2}+{q}^{2} \,.
\]

Nevertheless, to verify the existence of black hole Meissner effect in the presence of NUT parameter, we can consider the vanishing electric charge of the solutions above, i.e. $q=0$. Furthermore, we can perform a gauge transformation to the vector (\ref{An}) with $q=0$ by adding a constant one form
\be 
{\frac {4 b \left( {a}^{2}{b}^{2}{l}^{2}-{b}^{2}{l}^{4}+{a}^{2}+4{a}^{4}{b}^{2} \right) }{4{a}^{2}{b}^{4}{l}^{2}-4{b}^{4}{l}^{4}+8{a}^{2}{b}^{2}+16{a}^{4}{b}^{4}+1}} d\phi '
\ee  
that gives us the near horizon vector field
\be \label{Anq0}
A_\mu  dx^\mu   = {\left. {L\left( x \right)} \right|_{q = 0} } r'dt' + A_{\phi '} d\phi '\,,
\ee 
where
\be \label{App}
A_{\phi '} = \frac{8{a}^{2}b x}{\cal K} \left(4{b}^{4}{l}^{5}+4ax{b}^{4}{l}^{4}+4{b}^{2} \left( 3{a}^{2}{b}^{2}+1 \right)  {l}^{3}-4{a}^{3}x{b}^{4}{l}^{2}+ \left( 4{a}^{2}{b}^{2}+1 \right) l+ax-16{a}^{5}x{b}^{4}
 \right) 
\ee 
with
\[
{\cal K} = \left( a{x}^{2}+16{a}^{5}{b}^{4}+8{a}^{3}{b}^{2}+2lx-8{a}^{3}
{b}^{2}{x}^{2}+16{a}^{5}{b}^{4}{x}^{2}+4{a}^{3}{b}^{4}{l}^{2}-4a
{b}^{4}{l}^{4}+4{a}^{3}{b}^{4}{l}^{2}{x}^{2}\right.
\]
\[
\left. +a -4a{b}^{4}{l}^{4}{x}
^{2}-24{a}^{2}{b}^{4}{l}^{3}x-8{b}^{4}{l}^{5}x \right)  \left( 1+
16{a}^{4}{b}^{4}+8{a}^{2}{b}^{2}-4{b}^{4}{l}^{4}+4{a}^{2}{b}^{4}{l}^{2} \right)  \,.
\]

In the case of $q=0$, the constant $k$ in the near horizon metric (\ref{nhmetric}) vanishes for the magnetic field parameter $b=H^{-1/2}$ where $H = 2\sqrt {23l^4  - 19a^2 l^2  + 4a^4 } $ and the near horizon geometry becomes static. Accordingly, in this static limit, the vector component (\ref{App}) can be expressed as
\be \label{Anq0b0}
\left. {A_{\phi '}} \right|_{b = \frac{1}{{\sqrt H }}}  = \frac{{2\sqrt 2 a^2 lxH^{3/2} \left( {\zeta _1 H + \chi _1 } \right)}}{{\left( {\zeta _2 H + \chi _2 } \right)\left( {\zeta _3 H + \chi _3 } \right)}}
\ee 
with
\[
\chi_1 = 2al \left( 5{a}^{2} -6{l}^{2} \right) x + 2{a}^{4}-8{a}^{2}{l}^{2}+12{l}^{4}\,,
\]
\[
\zeta_1 = l^2 + a^2\,,
\]
\[
\chi_2 = 11{l}^{4}-9{a}^{2}{l}^{2}+4{a}^{4}\,,
\]
\[
\zeta_2 = 2a^2\,,
\]
\[
\chi_3 = a \left( 11{l}^{4}-9{a}^{2}{l}^{2}+4{a}^{4} \right) \left(1+x^2\right) + 2l \left( 2{a}^{4} -11{a}^{2}{l}^{2}+11{l}^{4} \right) \,,
\]
and
\[
\zeta_3 = 2a^3 \Delta_x\,.
\]
Obviously, the function (\ref{Anq0b0}) vanishes for $l=0$ which confirms the black hole Meissner effect reported in \cite{Bicak:2015lxa}. However, in the presence of NUT parameter $l$, the static limit of near horizon metric (\ref{nhmetric}) does not lead to the expulsion of axial component in the near horizon vector field. Therefore, we can infer that the black hole Meissner effect does not exist in the case of extremal  Melvin-Kerr-Taub-NUT black hole, and the conclusion in general should be true in the case of extremal Melvin-Kerr-Newman-Taub-NUT case. 

\begin{figure}[h]
\begin{center}
		\includegraphics[scale=0.5]{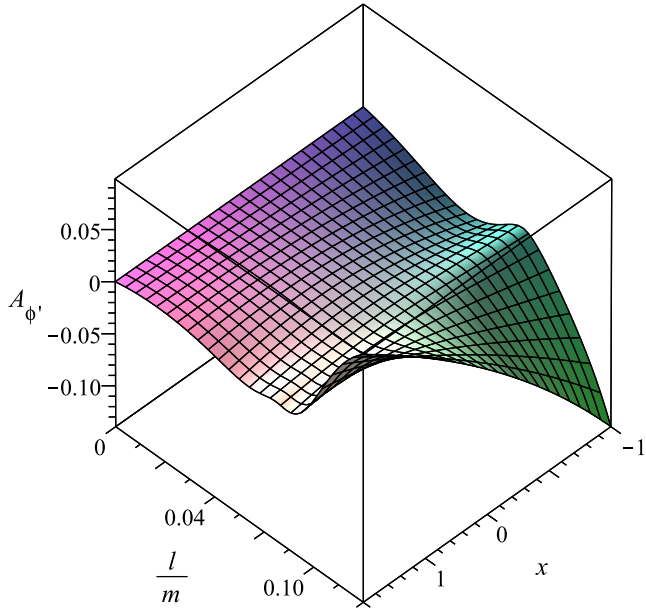}\caption{The magnitude of $A_{\phi '}$ evaluated at $b=H^{-1/2}$ and $a=0.1 m$. Note that the extremal condition $m^2 + l^2 = a^2$ has been applied.}\label{fig.Ap}
\end{center}
\end{figure}

\begin{figure}[h]
	\begin{center}
		\includegraphics[scale=0.5]{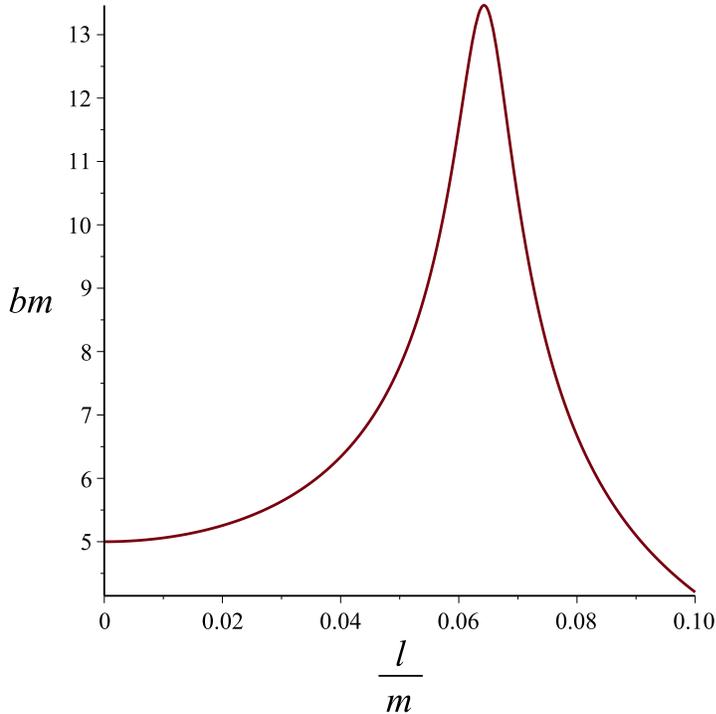}\caption{The corresponding $b=H^{-1/2}$ for $a=0.1 m$ that is used to produce the plot in fig. \ref{fig.Ap}. Note that in the range of $l = 0$ to $l = 0.1 n$, the value of $b=H^{-1/2}$ is real.}\label{fig.b}
	\end{center}
\end{figure}

To illustrate the dependence of $A_{\phi '}$ function with respect to the NUT parameter for the external magnetic field parameter $b=H^{-1/2}$, we provide figures \ref{fig.Ap} and \ref{fig.b}. From the plots, we can verify that the Meissner which occurs in the $l=0$ case ceases to exist in the presence of NUT parameter $l$. Indeed the solution $b=-H^{-1/2}$ also yields the vanishing of $k$ in the near horizon metric (\ref{nhmetric}), i.e. it leads to the static case, but it does not change the fact that black hole Meissner effect does not appear in the Melvin-Taub-NUT black hole. 

\section{Conclusions}
We explore correspondence between the MKNTN black holes, and the conformal field theory, by finding the agreement between the Cardy entropy and the entropy of the extremal MKNTN black holes. We show that the near horizon geometry of extremal MKNTN black hole possesses the  $SL\left(2,{\mathbb R}\right)\times U\left(1\right)$ symmetry. We consider the asymptotic symmetry group method, to find the central charge and the entropy of the dual CFT. We also use the alternative way of the stretched horizon method, to compute the extremal entropy of the MKNTN black hole. We find perfect agreement between the results of asymptotic symmetry group and the stretched horizon method. The results of this paper strongly supports the holographic dual for the extremal MKNTN black holes.

In the Kerr/CFT holography, we learn that Ernst magnetization changes the associated central charge and Frolov-Thorne temperature in the dual CFT formula for entropy. However, the Kerr/CFT holography is still valid in the magnetized spacetime provided it exists in the non-magnetized counterpart. Both the ASG and SHM methods for Kerr/CFT correspondence exhibit this property, as it has shown earlier in \cite{Siahaan:2015xia,Astorino:2015lca}. Note that, as it was also reported in \cite{Siahaan:2015xia} for the null NUT parameter case, the dual CFT describing a MKNTN black hole embedded in a strong magnetic field can be non-unitary for a negative value of central charge. Nevertheless, the entropy of an extremal MKNTN black hole is always positive even in the possible non-unitary system.

We explore the black hole Meissner effect as reported in \cite{Bicak:2015lxa} for the Melvin-Taub-NUT system. In this magnetized spacetime, the presence of the NUT parameter prevents the expulsion of the axial component of the near-horizon vector field at extremality. This is due to contributions of the NUT parameter to the axial component of the gauge field, which does not vanish at the static limit of an extreme black hole. Interestingly, in \cite{Bicak:2015lxa} it is demonstrated that the near-horizon limits of extremal Kerr-Newman and Melvin-Kerr-Newman black holes are mathematically equivalent. We aim to extend this result by showing the equivalence between the near-horizon limits of extremal Kerr-Newman-Taub-NUT and Melvin-Kerr-Newman-Taub-NUT spacetimes. However, the complexity of the Melvin-Kerr-Newman-Taub-NUT solution requires careful treatment for this investigation. Additionally, even in the absence of the NUT charge, the expressions for $\Omega_H$, $\Phi_H$, $Q$, and $J$ in the total mass of the magnetized spacetime are already quite lengthy. Finding an exact expression for the magnetic moment $\mu$ would be a challenging task. In \cite{Gibbons:2013dna}, an expression for $\mu$ up to linear order in $q$ is presented, whereas $\Omega_H$ and $\Phi_H$ are linear orders in $b$. We will address these issues in our future work.

\section*{Acknowledgement}

MG was supported by the Natural Sciences and Engineering Research Council of Canada. HMS was supported by the Lembaga Penelitian dan Pengabdian Kepada Masyarakat UNPAR under contract no. III/LPPM/2023-02/45-P.

\appendix

\section{Melvin-Kerr-Newman-Taub-NUT spacetime and its thermodynamics}
\label{sec:MKNTN}

Melvin-Kerr-Newman-Taub-NUT spacetime is obtained after applying Ernst transformation to the Kerr-Newman-Taub-NUT solution as the seed \cite{Ghezelbash:2021lcf}. The magnetization can be considered as a procedure used to generate magnetized black hole solutions by making use of the global $SU(2,1)$ symmetry group \cite{Gibbons:2013yq,Stephani:2003tm,Galtsov:2008zz}. As presented in \cite{Gibbons:2013yq}, the symmetry group is obtained after performing a Kaluza-Klein reduction of the four-dimensional Einstein-Maxwell action and dualizing the vector fields to scalars in three dimensions. After obtaining the corresponding reduced three-dimensional Lagrangian, adding Lagrange multipliers and eliminating some terms allow us to achieve a dualized Lagrangian which can be expressed as
\be 
{\cal L}_3  = R - {\rm{tr}}\left( {\cal M}^{ - 1} \partial {\cal M} \right)^2 \,.
\ee
This dualized Lagrangian is invariant under $SU(2,1)$, with
\be 
{\cal M} \to {\cal M}' = U^\dagger {\cal M} U
\ee 
where $U$  is a $SU(2,1)$ matrix with constant elements. Related to the Ernst magnetization with external magnetic parameter $b$, then $U$ is parameterized this variable.

In the Lewis-Papapetrou-Weyl (LPW) form
\be\label{metricmag} 
{\rm{d}}s^2  = - f^{ - 1} \rho ^2 {\rm{d}}t^2  + f^{ - 1}e^{2\gamma } \left( \frac{{{\rm{d}}r^2 }}{{\Delta _r }} + \frac{{{\rm{d}}x^2 }}{{\Delta _x }} \right)  + f\left( {\omega {\rm{d}}t-{\rm{d}}\phi } \right)^2,
\ee
with $\Delta_r = r^2-2mr+a^2+q^2-l^2$ and $\Delta_x = 1-x^2$, the MKNTN solution can be expressed with the metric functions
\be 
f = \frac{{\sum\limits_{k = 0}^4 {{c_k}{x^k}} }}{{\sum\limits_{j = 0}^6 {{d_j}{x^j}} }},\label{ff}
\ee 
where
\[
c_4 = -a^2 \Delta_r,
\]
\[
c_3 = -4la \Delta_r,
\]
\[
c_2 = 3{l}^{4}+ \left( 8rm-8{a}^{2}-4{q}^{2}-6{r}^{2} \right) {l}^{2}+{a}^{4}-4{a}^{2}mr+2{a}^{2}{q}^{2}-{r}^{4},
\]
\[
c_1 = 4la \Delta_r,
\]
\[
c_0 = 3{a}^{2}{l}^{2}+2{a}^{2}mr-{a}^{2}{q}^{2}+{a}^{2}{r}^{2}+{l}^{4}+2{l}^{2}{r}^{2}+{r}^{4},
\]
\[
d_6 = a^2 b^4 \Delta_r^2,
\]
\[
d_5 = 6 a b^4 l\Delta_r^2 ,
\]
\[
d_4 = -b\left\{  2{a}^{6}{b}^{2}-21{a}^{4}{b}^{2}{l}^{2}-4{a}^{4}{b}^{2}{m}^{2}-
12{a}^{4}{b}^{2}mr+6{a}^{4}{b}^{2}{q}^{2}+3{a}^{4}{b}^{2}{r}^{2}
+28{a}^{2}{b}^{2}{l}^{4}\right.
\]
\[
-8{a}^{2}{b}^{2}{l}^{2}{m}^{2}+84{a}^{2}
{b}^{2}{l}^{2}mr-32{b}^{2}{a}^{2}{l}^{2}{q}^{2}-36{a}^{2}{b}^{2}{l}^{2}{r}^{2}+24{a}^{2}{b}^{2}{m}^{2}{r}^{2}
\]
\[
-24{a}^{2}{b}^{2}m{q}^{2}r-12{a}^{2}{b}^{2}m{r}^{3}+4{a}^{2}{b}^{2}{q}^{4}+8{a}^{2}{b}^{2}{q}^{2}{r}^{2}-9{b}^{2}{l}^{6}-4{b}^{2}{l}^{4}{m}^{2}-24{b}^{2}{l}^{4}mr
\]
\[
+18{b}^{2}{l}^{4}{q}^{2}+9{b}^{2}{l}^{4}{r}^{2}-36{b}^{2}{l}^{2}{m}^{2}{r}^{2}+32{b}^{2}{l}^{2}m{q}^{2}r+40{b}^{2}{l}^{2}m{r}^{3}-9{b}^{2}{l}^{2}{q}^{4}
\]
\[
-12{b}^{2}{l}^{2}{q}^{2}{r}^{2}-15{b}^{2}{l}^{2}{r}^{4}-{b}^{2}{q}^{4}{r}^{2}+2{b}^{2}{q}^{2}{r}^{4}-{b}^{2}{r}^{6}-4{a}^{3}bqr+4ab{l}^{2}qr
\]
\[
\left. +8abmq{r}^{2}-4ab{q}^{3}r-4abq{r}^{3}+2{a}^{4}-2{a}^{2}{l}^{2}-4{a}^{2}mr+2{a}^{2}{q}^{2}+2{a}^{2}{r}^{2} \right\},
\]
\[
d_3 = -4b^2 l\left\{ 3{a}^{5}{b}^{2}-10{a}^{3}{b}^{2}{l}^{2}+2{a}^{3}{b}^{2}{m}^{2}-
20{a}^{3}{b}^{2}mr+6{a}^{3}{b}^{2}{q}^{2} +6{a}^{3}{b}^{2}{r}^{2}\right.
\]
\[
+7a{b}^{2}{l}^{4}+2a{b}^{2}{l}^{2}{m}^{2}+16a{b}^{2}{l}^{2}mr-10
a{b}^{2}{l}^{2}{q}^{2}-2a{b}^{2}{l}^{2}{r}^{2}+18a{b}^{2}{m}^{2}{r}^{2}
\]
\[
-14a{b}^{2}m{q}^{2}r-12a{b}^{2}m{r}^{3}+3a{b}^{2}{q}^{4}+
2a{b}^{2}{q}^{2}{r}^{2}+3a{b}^{2}{r}^{4}+2{a}^{2}bmq-4{a}^{2}b
qr
\]
\[
\left. +2b{l}^{2}mq+6bmq{r}^{2}-2b{q}^{3}r-4bq{r}^{3}+2{a}^{3}-2
a{l}^{2}-4amr+2a{q}^{2}+2a{r}^{2} \right\},
\]
\[
d_2 = {a}^{6}{b}^{4}-28{a}^{4}{b}^{4}{l}^{2}+8{a}^{4}{b}^{4}{m}^{2}-12
{a}^{4}{b}^{4}mr+4{a}^{4}{b}^{4}{q}^{2}+37{a}^{2}{b}^{4}{l}^{4}+12
{a}^{2}{b}^{4}{l}^{2}{m}^{2}
\]
\[
+84{a}^{2}{b}^{4}{l}^{2}mr-38{a}^{2}
{b}^{4}{l}^{2}{q}^{2}-18{a}^{2}{b}^{4}{l}^{2}{r}^{2}+36{a}^{2}{b}^
{4}{m}^{2}{r}^{2}-20{a}^{2}{b}^{4}m{q}^{2}r
\]
\[
-12{a}^{2}{b}^{4}m{r}^{3}+4{a}^{2}{b}^{4}{q}^{4}+6{a}^{2}{b}^{4}{q}^{2}{r}^{2}-3{a}^{2}
{b}^{4}{r}^{4}+6{b}^{4}{l}^{6}+24{b}^{4}{l}^{4}mr-6{b}^{4}{l}^{4
}{q}^{2}
\]
\[
-30{b}^{4}{l}^{4}{r}^{2}-8{b}^{4}{l}^{2}m{r}^{3}+12{b}^{4}{l}^{2}{q}^{2}{r}^{2}-6{b}^{4}{l}^{2}{r}^{4}+2{b}^{4}{q}^{2}{r}^
{4}-2{b}^{4}{r}^{6}+8{a}^{3}{b}^{3}mq
\]
\[
-8{a}^{3}{b}^{3}qr+16a{b}
^{3}{l}^{2}mq-8a{b}^{3}{l}^{2}qr+24a{b}^{3}mq{r}^{2}-4a{b}^{3}{q
}^{3}r-8a{b}^{3}q{r}^{3}+2{a}^{4}{b}^{2}
\]
\[
-16{a}^{2}{b}^{2}{l}^{2}
-8{a}^{2}{b}^{2}mr+4{a}^{2}{b}^{2}{q}^{2}+6{b}^{2}{l}^{4}+16{b}^{2}{l}^{2}mr-2{b}^{2}{l}^{2}{q}^{2}-12{b}^{2}{l}^{2}{r}^{2}
\]
\[
+6{b}^{2}{q}^{2}{r}^{2}-2{b}^{2}{r}^{4}-4abqr+{a}^{2},
\]
\[
d_1 = 2l\left\{ 3{a}^{5}{b}^{4}-14{a}^{3}{b}^{4}{l}^{2}-4{a}^{3}{b}^{4}{m}^{2}-
12{a}^{3}{b}^{4}mr+6{a}^{3}{b}^{4}{q}^{2}+6{a}^{3}{b}^{4}{r}^{2}
-5a{b}^{4}{l}^{4} \right.
\]
\[
-12a{b}^{4}{l}^{2}mr+2a{b}^{4}{l}^{2}{q}^{2}+18
a{b}^{4}{l}^{2}{r}^{2}+4a{b}^{4}m{r}^{3}-2a{b}^{4}{q}^{2}{r}^{2}
+3a{b}^{4}{r}^{4}-8{a}^{2}{b}^{3}mq
\]
\[
\left. -8{b}^{3}{l}^{2}qr+4{b}^{2}
{a}^{3}-4a{b}^{2}{l}^{2}-8a{b}^{2}mr-2a{b}^{2}{q}^{2}+4a{b}^{2
}{r}^{2}-4bqr+a \right\},
\]
\[
d_0 = 9{a}^{4}{b}^{4}{l}^{2}+4{a}^{4}{b}^{4}{m}^{2}+4{a}^{4}{b}^{4}mr+
{a}^{4}{b}^{4}{r}^{2}+6{a}^{2}{b}^{4}{l}^{4}+12{a}^{2}{b}^{4}{l}^{2}mr+4{a}^{2}{b}^{4}m{r}^{3}
\]
\[
+2{a}^{2}{b}^{4}{r}^{4}+{b}^{4}{l}^{6}
+7{b}^{4}{l}^{4}{r}^{2}+7{b}^{4}{l}^{2}{r}^{4}+{b}^{4}{r}^{6}+8{a}^{3}{b}^{3}mq+4{a}^{3}{b}^{3}qr+12a{b}^{3}{l}^{2}qr
\]
\[
+4a{b}^{3}q
{r}^{3}+6{a}^{2}{b}^{2}{l}^{2}+4{a}^{2}{b}^{2}mr+4{a}^{2}{b}^{2}
{q}^{2}+2{a}^{2}{b}^{2}{r}^{2}+2{b}^{2}{l}^{4}+4{b}^{2}{l}^{2}{r}^{2}
\]
\be
+2{b}^{2}{r}^{4}+4abqr+{l}^{2}+{r}^{2},
\ee
and 
\be 
\omega = -\frac{{\sum\limits_{k = 0}^6 {{{\tilde c}_k}{x^k}} }}{{\sum\limits_{j = 0}^4 {{c_j}{x^j}} }},\label{oomega}
\ee 
where
\[
{\tilde c}_6 = a{b}^{4} \Delta_r \left( 
2{a}^{2}{l}^{2}-4{m}^{2}{a}^{2}+6{a}^{2}mr-{a}^{2}{q}^{2}-2{l}
^{4}-4{l}^{2}{m}^{2}+6{l}^{2}mr+3{l}^{2}{q}^{2} \right. 
\]
\[
\left. -6{l}^{2}{r}^{2
}-2m{r}^{3}-{q}^{4}+3{q}^{2}{r}^{2} \right),
\]
\[
{\tilde c}_5 = -2{b}^{3}l \Delta_r 
\left( 3{a}^{4}b-6{a}^{2}b{l}^{2}+4{a}^{2}b{q}^{2}-6{a}^{2}b{
	r}^{2}+3b{l}^{4}+4b{l}^{2}{m}^{2} -4b{l}^{2}mr-4b{l}^{2}{q}^{2}
+6b{l}^{2}{r}^{2} \right.
\]
\[
\left. +4bm{r}^{3}+b{q}^{4}-4b{q}^{2}{r}^{2}-b{r}^{4}-
4qma+4qar \right),
\]
\[
{\tilde c}_4 = -{b}^{2} \Delta_r  \left( 
32{a}^{3}{b}^{2}{l}^{2}+8{a}^{3}{b}^{2}{m}^{2}-6{a}^{3}{b}^{2}mr
-30a{b}^{2}{l}^{4}-20a{b}^{2}{l}^{2}{m}^{2}+18a{b}^{2}{l}^{2}mr+
25{q}^{2}{l}^{2}{b}^{2}a \right.
\]
\[
-42a{b}^{2}{l}^{2}{r}^{2}-14a{b}^{2}m{r}
^{3}-a{b}^{2}{q}^{4}+9{r}^{2}a{b}^{2}{q}^{2}+16{a}^{2}bmq-12r{a}
^{2}bq-16b{l}^{2}mq
\]
\[
\left. +12rb{l}^{2}q-4{r}^{3}bq+6a{q}^{2} \right),
\]
\[
{\tilde c}_3 = 4{b}^{2}l \Delta_r 
\left( {a}^{4}{b}^{2}-20{a}^{2}{b}^{2}{l}^{2}-8{a}^{2}{b}^{2}{m}^
{2}+6{a}^{2}{b}^{2}mr+3{a}^{2}{b}^{2}{q}^{2}-6{a}^{2}{b}^{2}{r}^
{2}+{b}^{2}{l}^{4}-2{b}^{2}{l}^{2}mr \right.
\]
\[
\left. -2{b}^{2}{l}^{2}{q}^{2}+6{b}
^{2}{l}^{2}{r}^{2}+2{b}^{2}m{r}^{3}-2{b}^{2}{q}^{2}{r}^{2}+{b}^{2}
{r}^{4}-10abmq+8abqr-3{q}^{2} \right),
\]
\[
{\tilde c}_2 = 24{a}^{5}{b}^{4}{l}^{2}+8{a}^{5}{b}^{4}{m}^{2}-14r{a}^{5}{b}^{4}
m-92{a}^{3}{b}^{4}{l}^{4}-16{a}^{3}{b}^{4}{l}^{2}{m}^{2}-36r{a}^
{3}{b}^{4}{l}^{2}m+46{a}^{3}{b}^{4}{l}^{2}{q}^{2}
\]
\[
-12{r}^{2}{a}^{3}
{b}^{4}{l}^{2}-24r{a}^{3}{b}^{4}{m}^{3}+12{a}^{3}{b}^{4}{m}^{2}{q}
^{2}+48{a}^{3}{b}^{4}{m}^{2}{r}^{2}-20{a}^{3}{b}^{4}m{q}^{2}r-20
{r}^{3}{b}^{4}{a}^{3}m
\]
\[
+6{r}^{2}{a}^{3}{b}^{4}{q}^{2}+28a{b}^{4}{l}
^{6}+58ra{b}^{4}{l}^{4}m-46a{b}^{4}{l}^{4}{q}^{2}-24{r}^{2}a{b}^
{4}{l}^{2}{m}^{2}-8ra{b}^{4}{l}^{2}m{q}^{2}
\]
\[
+108a{b}^{4}{l}^{2}m{r}^{3}+10a{b}^{4}{l}^{2}{q}^{4}-36{r}^{2}a{b}^{4}{l}^{2}{q}^{2}-36
a{b}^{4}{l}^{2}{r}^{4}+24{b}^{4}a{m}^{2}{r}^{4}-24{r}^{3}{b}^{4}am
{q}^{2}
\]
\[
-6a{b}^{4}m{r}^{5}+6{r}^{2}a{b}^{4}{q}^{4}+6{b}^{4}a{q}^{2}{r}^{4}+8{a}^{4}{b}^{3}mq-16r{a}^{4}{b}^{3}q-24{a}^{2}{b}^{3}{l}^{2}mq+32r{a}^{2}{b}^{3}{l}^{2}q
\]
\[
-32r{a}^{2}{b}^{3}{m}^{2}q+16{a}^{2}{b}^{3}m{q}^{3}+48{a}^{2}{b}^{3}mq{r}^{2}-20r{a}^{2}{b}^{3}{q}^{3}-16{r}^{3}{b}^{3}{a}^{2}q-16r{b}^{3}{l}^{4}q
\]
\[
-24{r}^{2}{b}^{3}{l}^{2}mq+12r{b}^{3}{l}^{2}{q}^{3}+16{r}^{3}{b}^{3}{l}^{2}q+8
{b}^{3}mq{r}^{4}-4{r}^{3}{b}^{3}{q}^{3}-12{q}^{2}{l}^{2}{b}^{2}a
\]
\[
-12ra{b}^{2}m{q}^{2}+6a{b}^{2}{q}^{4}-4r{a}^{2}bq-4rb{l}^{2}q-4
{r}^{3}bq+2a{l}^{2}+2amr-a{q}^{2},
\]
\[
{\tilde c}_1 = 2l \Delta_r  \left( {a}^
{4}{b}^{4}+18{a}^{2}{b}^{4}{l}^{2}-12{a}^{2}{b}^{4}mr-2{a}^{2}{b
}^{4}{q}^{2}+6{a}^{2}{r}^{2}{b}^{4}+{b}^{4}{l}^{4}\right.
\]
\[
\left. -6{b}^{4}{l}^{2}
{r}^{2}-3{b}^{4}{r}^{4}-12a{b}^{3}qr-1 \right),
\]
\[
{\tilde c}_0 = 6{a}^{5}{b}^{4}{l}^{2}+4{a}^{5}{b}^{4}{m}^{2}+2r{a}^{5}{b}^{4}m+
{a}^{5}{b}^{4}{q}^{2}+26{a}^{3}{b}^{4}{l}^{4}+4{a}^{3}{b}^{4}{l}^{2}{m}^{2}-7{a}^{3}{b}^{4}{l}^{2}{q}^{2}+6{r}^{2}{a}^{3}{b}^{4}{l}^{2}
\]
\[
-12{a}^{3}{b}^{4}{m}^{2}{r}^{2}+6{a}^{3}{b}^{4}m{q}^{2}r-4{r}
^{3}{b}^{4}{a}^{3}m+{a}^{3}{b}^{4}{q}^{4}+{r}^{2}{a}^{3}{b}^{4}{q}^{2}
+8a{b}^{4}{l}^{6}+2ra{b}^{4}{l}^{4}m-12a{b}^{4}{l}^{2}m{r}^{3}
\]
\[
-6a{b}^{4}m{r}^{5}+8{a}^{4}{b}^{3}m
q+4r{a}^{4}{b}^{3}q+8{a}^{2}{b}^{3}{l}^{2}mq-8r{a}^{2}{b}^{3}{l}^{2}q-8{a}^{2}{b}^{3}mq{r}^{2}+8r{a}^{2}{b}^{3}{q}^{3}+4r{b}^{3}
{l}^{4}q-4{b}^{3}q{r}^{5}
\]
\be
+6{a}^{3}{b}^{2}{q}^{2}+6{q}^{2}{l}^{2}
{b}^{2}a+6{r}^{2}a{b}^{2}{q}^{2}+4r{a}^{2}bq+4rb{l}^{2}q+4{r}^{3}bq-2a{l}^{2}-2amr+a{q}^{2}+4
{r}^{2}a{b}^{4}{l}^{2}{q}^{2}.
\ee
Moreover the metric function $\gamma$ is given by
\[
{e^{2\gamma }} = \Delta_x r^4 +\left[2{l}^{2}-{a}^{2}{x}^{4}-4al{x}^{3}-6{l}^{2}{x}^{2}+4alx+{a}^{2}
\right] r^2 + 2m \left[ a{x}^{2}+2lx-a \right] ^{2} r
\]
\[ 
-{a}^{2} \left[ {a}^{2}-{l}^{2}+{q}^{2} \right] x^4  +\left[{a}^{4}-8{a}^{2}{l}^{2}+2{a}^{2}{q}^{2}+3{l}^{4}-4{l}^{2}{q}^{2}\right] x^2 
\] 
\be\label{ggama}
-4al \left[ {a}^{2}-{l}^{2}+{q}^{2} \right] x^3 +4 al \left[ {a}^{2}-{l}^{2}+{q}^{2} \right] x +3{a}^{2}{l}^{2}-{a}^{2}{q}^{2}+{l}^{4},
\ee
and
\be
\rho^2 = \Delta_x \Delta_r.\label{rrho}
\ee
The gauge field $A_\mu$ for the MKNTN solution is given explicitly in \cite{Ghezelbash:2021lcf}. However, we do not repeat to present it here, due to its lengthy expression. In equations above, $m$, $q$, $l$, $a$, and $b$, are the parameters of mass, electric charge, NUT charge, rotation, and magnetic field, respectively.

The MKNTN solution solve the equations of motion with the action can be written as
\be 
S_{\rm EM} = \int  d^4 x \sqrt{-g}\left\{ R - \frac{1}{4} F_{\mu\nu} F^{\mu\nu} \right\}\,.
\ee 
It is known that there exist the electric-magnetic duality for the solutions in this theory, namely $E_k \to B_k$ and $B_k \to -E_k$, where $E_k$ and $B_k$ are the component of electric and magnetic fields in an orthonormal frame, respectively. Explicitly, we can write $E_k = F_{0k}$ and $B_k = {\tilde F}_{0k}$ where ${\tilde F}_{\mu\nu} = \tfrac{1}{2} \varepsilon_{\mu\nu \alpha\beta} F^{\alpha\beta}$ is the dual field strength tensor. Obviously, this duality holds for both of seed and magnetized solutions, i.e. the KNTN and MKNTN cases, since they are dictated by the same action above. In general, the electric-magnetic duality is a $U(1)$ rotation \cite{Gibbons:2013yq,Gibbons:1995cv}
\be \label{A.11}
F_{\mu\nu} \to \cos\alpha F_{\mu\nu} + \sin\alpha {\tilde F}_{\mu\nu}
\ee 
where the $Z_2$ symmetry mentioned above is a special case. It should be noted that equation \ref{A.11} does not represent an Ernst magnetization.

Interestingly, the area of the MKNTN black hole is given by \cite{Ghezelbash:2021lcf}
\be \label{area}
{\cal A}_H = 4\pi \left(2 l^2 + 2m r_+ -q^2\right),
\ee 
which is the same as the area of the Kerr-Newman-Taub-NUT (KNTN) black hole. The resemblance between corresponding quantities for the MKNTN and KNTN black holes occurs also for the Hawking temperature,
\be
T_H=\frac{\kappa}{2\pi}\,,
\ee
where the surface gravity $\kappa$ is given by
\be
\kappa=\frac{1}{2r_+}\frac{r_+^2+l^2-a^2-q^2}{r_+^2+l^2+a^2}.\label{kappa}
\ee

Studying the thermodynamics of the black holes with the NUT parameter, is an active research subject \cite{BallonBordo:2019vrn}. In  \cite{Ghezelbash:2021lcf}, the authors established the Smarr relation for the MKNTN spacetime without a term containing the magnetic moment of the system. This is because magnetic field is not considered as an additional thermodynamic variable in the first law. An alternative to this approach can be done, namely magnetic field is considered to be a new thermodynamical variable \cite{Gibbons:2013dna,Astorino:2016hls}, which leads to the Smarr formula 
\be\
{\cal M}=\Phi_H Q_0+2\Omega_H {J}+\frac{\kappa}{4\pi}{\cal A}_H+2\psi_+ N_++2\psi_- N_- + \mu b\,, \label{MassTotal}
\ee 
where the total mass ${\cal M}$ obeys the first thermodynamics law
\be \label{dMtot}
{\delta }{\cal M}=\Phi_H {\delta }Q_0+ \Omega_H {\delta }{J}+T_H {\delta }S+ \psi_+ {\delta }N_++ \psi_- {\delta }N_- -\mu \delta b\,.
\ee

The term $\mu b$ introduced above denotes the system's "magnetization", with $\mu$ representing the magnetic moment. To differentiate them from those used in \cite{Ghezelbash:2021lcf}, we use distinct notations for the charge and angular momentum quantities, which are defined as follows:
\be 
Q_0  = \left. {\frac{{q_0 }}{{\sum\limits_{i = 0}^4 {s_i b^i } }}} \right|_{r = r_ +  } \,
\ee
and
\be 
{J}  = \left. {\frac{{j_0 }}{{\sum\limits_{i = 0}^8 {k_i b^i } }}} \right|_{r = r_ +  } \,
\ee 
where all the coefficients $q_i$, $s_i$, and $k_i$ are those from \cite{Ghezelbash:2021lcf}. Consequently, the magnetization term can be expressed as:
\be 
\mu b = \Phi _H \left( {Q - Q_0 } \right) + 2\Omega _H \left( {{\cal J} - {J} } \right)\,.
\ee 
In equations above, $\Omega_H$ is the angular velocity of the horizon $\Omega_H$, $\Phi_H$ is the Coulomb potential, and $Q$ is total charge. The horizon area is ${\cal A}_H$, and the surface gravity is denoted by $\kappa$. The Misner potentials are given by $\psi_{\pm}$, and the corresponding Misner charges are $N_{\pm}$. 

It is understood that the entropy of MKNTN spacetime depends on the incorporating charges, i.e. ${\cal S}\left(Q_0,{{J}}, N_+, N_-,b\right)$. Accordingly, we can define the following chemical potentials \cite{Compere:2012jk}
\[ 
\frac{1}{{{T_\phi }}} = {\left( {\frac{{\partial {{\cal S}_{{\rm{ext}}}}}}{{\partial {{J}}}}} \right)_{b,Q_0,{N_ + },{N_ - }}}~,~\frac{1}{{{T_q}}} = {\left( {\frac{{\partial {{\cal S}_{{\rm{ext}}}}}}{{\partial Q_0}}} \right)_{b,{{J}},{N_ + },{N_ - }}}~,
\]
\be \label{chem.pot}
~\frac{1}{{{T_ {\pm} }}} = {\left( {\frac{{\partial {{\cal S}_{{\rm{ext}}}}}}{{\partial {J}}}} \right)_{b,{{J}},Q_0,{N_{\mp} }}}~,~\frac{1}{{{T_b}}} = {\left( {\frac{{\partial {{\cal S}_{{\rm{ext}}}}}}{{\partial Q_0}}} \right)_{Q_0,{{J}},{N_ + },{N_ - }}}\,,
\ee 
where ${\cal S}_{\rm ext}$ is the extremal entropy which fulfills the balance equation
\be \label{dSext}
\delta {{\cal S}_{{\rm{ext}}}} = \frac{1}{{{T_\phi }}}\delta {{J}} + \frac{1}{{{T_q}}}\delta Q_0 + \frac{1}{{{T_ + }}}\delta {N_ + } + \frac{1}{{{T_ - }}}\delta {N_ - }+\frac{1}{{{T_b}}}\delta b\,.
\ee 
Recall that the Hawking temperature $T_H$ vanishes at extremality. It implies 
\be \label{ThdSext}
{T_H}\delta S_{\rm ext} = \delta M - \left( {\Omega _H^{{\rm{ext}}}\delta {J} + \Phi _H^{{\rm{ext}}}\delta Q_0 + {\Psi _ + }\delta {N_ + } + {\Psi _ - }\delta {N_ - }}-\mu \delta b \right) = 0\,,
\ee 
which can be obtained based on eq. (\ref{dMtot}). Note that the results (\ref{chem.pot}), (\ref{dSext}), and (\ref{ThdSext}) are useful to obtain the Frolov-Thorne temperature (\ref{Tp}) near the extremal MKTN black hole in the next section.

\section{$L\left(x\right)$ function in the gauge field (\ref{nhA})}\label{app.Lx}

The $L\left(x\right)$ function, which appear in equation (\ref{nhA}), can be expressed as
\be 
L\left( x \right) = \frac{{\sum\limits_{j = 0}^4 {{{\lambda}_j}{x^j}} }}{{\sum\limits_{k = 0}^4 {{{\tilde {\lambda}}_k}{x^k}} }}\,,
\ee 
where
\[
{\lambda}_4 = \left( 32{a}^{6}{b}^{6}{m}^{2}q-16{a}^{6}{b}^{6}{q}^{3}-32{a}^{4}{b}^{6}{m}^{4}q+60{a}^{4}{b}^{6}{m}^{2}{q}^{3}-24{a}^{4}{b}^{6}{q}^{5}-28{a}^{2}{b}^{6}{m}^{4}{q}^{3}+28{a}^{2}{b}^{6}{m}^{2}{q}^{5}  \right.
\]
\[
+128{a}^{7}{b}^{5}m-80{a}^{5}{b}^{5}{m}
^{3}+176{a}^{5}{b}^{5}m{q}^{2}+16{a}^{3}{b}^{5}{m}^{5}-16{a}^{3}
{b}^{5}{m}^{3}{q}^{2}+4m{a}^{3}{b}^{5}{q}^{4}-16a{b}^{5}{m}^{5}{q}
^{2}+48a{b}^{5}{m}^{3}{q}^{4}
\]
\[
-48{a}^{6}{b}^{4}q+180{a}^{4}{b}^{4}{m}^{2}q-80{a}^{4}{b}^{4}{q}^{3}-52{a}^{2}{b
}^{4}{m}^{4}q+132{a}^{2}{b}^{4}{m}^{2}{q}^{3}-29{a}^{2}{b}^{4}{q}^
{5}+16{b}^{4}{m}^{4}{q}^{3}-24{b}^{4}{m}^{2}{q}^{5}+4{b}^{4}{q}^
{7}
\]
\[
-48m{a}^{5}{b}^{3}+16{a}^{3}{b}^{3}{m}^{3}-56{a}^{3}{b}^{3}m{
	q}^{2}+48a{b}^{3}{m}^{3}{q}^{2}-16ma{b}^{3}{q}^{4}+8{a}^{4}{b}^{2}q-32{a}^{2}{b}^{2}{m}^{2}q+5{a}^{2}{b}^{2}{q}^{3}+8{b}^{2}{m}^
{2}{q}^{3}-4{b}^{2}{q}^{5}
\]
\[
\left. -44ma{b}^{5}{q}^{6}-9{a}^{2}{b}^{6}{q}^{7}+4{a}^{3}bm -4abm{q}^{2}+{a}^{2}q 
\right) \left( 16{a}^{6}{b}^{4}+4{a}^{4}{b}^{4}{m}^{2}+24{a}^{
	4}{b}^{4}{q}^{2}-4{a}^{2}{b}^{4}{m}^{4}+4{a}^{2}{b}^{4}{m}^{2}{q}^
{2}\right.
\]
\[
\left.   +9{a}^{2}{b}^{4}{q}^{4}+8{a}^{3}{b}^{3}mq +8a{b}^{3}{m}^{3}q+4a{b}^{3}m{q}^{3}-8{a}^{4}{b}^{2}-2{a}^{2}{b}^{2}{q}^{2}+4{b}^{
	2}{q}^{4}-4abmq+{a}^{2} \right),
\]
\[
{\lambda}_3 = -4\sqrt {{a}^{2}-{m}^{2}+{q}^{2}} \left( 128{a}^{9}{b}^{10}{m}^{2}{q}^{3}-256{a}^{11}{b}^{10}{q}^{
	3}-704{a}^{9}{b}^{10}{m}^{4}q-640
{a}^{9}{b}^{10}{q}^{5} +496{a}^{7}{b}^{10}{m}^{6}q\right.
\]
\[
-1488{a}^{7}{b}
^{10}{m}^{4}{q}^{3}+240{a}^{7}{b}^{10}{m}^{2}{q}^{5}-576{a}^{7}{b}
^{10}{q}^{7}-32{a}^{5}{b}^{10}{m}^{8}q+768{a}^{5}{b}^{10}{m}^{6}{q}^{3}-1032{a}^{5}{b}^{10}{m}^{4}{q}^{5}+144{a}^{5}{b}^{10}{m}^{2}{q}^{7}
\]
\[
-216{a}^{5}{b}^{10}{q}^{9}-16{a}^{3}{b}^{10}{m}^{10}q-48{a
}^{3}{b}^{10}{m}^{8}{q}^{3}+296{a}^{3}{b}^{10}{m}^{6}{q}^{5}-232{a
}^{3}{b}^{10}{m}^{4}{q}^{7}+27{a}^{3}{b}^{10}{m}^{2}{q}^{9}-27{a}^
{3}{b}^{10}{q}^{11}
\]
\[
+1024{a}^{12}{b}^{9}m-1920{a}^{10}{b}^{9}{m}^{3
}+2048{a}^{10}{b}^{9}m{q}^{2}+1184{a}^{8}{b}^{9}{m}^{5}-3264{a}^
{8}{b}^{9}{m}^{3}{q}^{2}+480{a}^{8}{b}^{9}m{q}^{4}-320{a}^{6}{b}^{
	9}{m}^{7}
\]
\[
+544{a}^{6}{b}^{9}{m}^{5}{q}^{2}-784{a}^{6}{b}^{9}{m}^{3}
{q}^{4}-1200{a}^{6}{b}^{9}m{q}^{6}+32{a}^{4}{b}^{9}{m}^{9}+160{a
}^{4}{b}^{9}{m}^{7}{q}^{2}-1152{a}^{4}{b}^{9}{m}^{5}{q}^{4}+1032{a
}^{4}{b}^{9}{m}^{3}{q}^{6}
\]
\[
-774{a}^{4}{b}^{9}m{q}^{8}+336{a}^{2}{b}
^{9}{m}^{7}{q}^{4}-672{a}^{2}{b}^{9}{m}^{5}{q}^{6}+444{a}^{2}{b}^{
	9}{m}^{3}{q}^{8}-108{a}^{2}{b}^{9}m{q}^{10}-768{a}^{11}{b}^{8}q+
3072{a}^{9}{b}^{8}{m}^{2}q
\]
\[
-1920{a}^{9}{b}^{8}{q}^{3}-3408{a}^{7}
{b}^{8}{m}^{4}q+5376{a}^{7}{b}^{8}{m}^{2}{q}^{3}-1728{a}^{7}{b}^{8}{q}^{5}+1248{a}^{5}{b}^{8}{m}^{6}q-4368{a}^{5}{b}^{8}{m}^{4}{q}^{3}+2232{a}^{5}{b}^{8}{m}^{2}{q}^{5}
\]
\[
-648{a}^{5}{b}^{8}{q}^{7}-144
{a}^{3}{b}^{8}{m}^{8}q+432{a}^{3}{b}^{8}{m}^{6}{q}^{3}-648{a}^{3}{b}^{8}{m}^{4}{q}^{5}-396{a}^{3}{b}^{8}{m}^{2}{q}^{7}-81{a}^{3}{b}^{8}{q}^{9}+96a{b}^{8}{m}^{8}{q}^{3}-384a{b}^{8}{m}^{6}{q}^{5}
\]
\[
+552
a{b}^{8}{m}^{4}{q}^{7}-264a{b}^{8}{m}^{2}{q}^{9}-256{a}^{10}{b}^
{7}m+640{a}^{8}{b}^{7}{m}^{3}-768{a}^{8}{b}^{7}m{q}^{2}-288{a}^{6}{b}^{7}{m}^{5}+2736{a}^{6}{b}^{7}{m}^{3}{q}^{2}-1184{a}^{6}{b}^{7}m{q}^{4}
\]
\[
+32{a}^{4}{b}^{7}{m}^{7}-1792{a}^{4}{b}^{7}{m}^{5}{q}^{2}+2848{a}^{4}{b}^{7}{m}^{3}{q}^{4}-1092{a}^{4}{b}^{7}m{q}^{6}+240
{a}^{2}{b}^{7}{m}^{7}{q}^{2}-1144{a}^{2}{b}^{7}{m}^{5}{q}^{4}+772
{a}^{2}{b}^{7}{m}^{3}{q}^{6}
\]
\[
-390{a}^{2}{b}^{7}m{q}^{8}-64{b}^{7}
{m}^{7}{q}^{4}+160{b}^{7}{m}^{5}{q}^{6}-112{b}^{7}{m}^{3}{q}^{8}+
16{b}^{7}m{q}^{10}+256{a}^{9}{b}^{6}q-432{a}^{7}{b}^{6}{m}^{2}q+
128{a}^{7}{b}^{6}{q}^{3}
\]
\[
+632{a}^{5}{b}^{6}{m}^{4}q+112{a}^{5}{b}
^{6}{m}^{2}{q}^{3}-360{a}^{5}{b}^{6}{q}^{5}-104{a}^{3}{b}^{6}{m}^{6}q+600{a}^{3}{b}^{6}{m}^{4}{q}^{3}+146{a}^{3}{b}^{6}{m}^{2}{q}^{5}-242{a}^{3}{b}^{6}{q}^{7}-208a{b}^{6}{m}^{6}{q}^{3}
\]
\[
+288a{b}^{6}
{m}^{4}{q}^{5}-252a{b}^{6}{m}^{2}{q}^{7}-96{a}^{8}{b}^{5}m-144{a}^{6}{b}^{5}m{q}^{2}+144{a}^{4}{b}^{5}{m}^{3}{q}^{2}-168{a}^{4}{b}
^{5}m{q}^{4}+72{a}^{2}{b}^{5}{m}^{5}{q}^{2}+348{a}^{2}{b}^{5}{m}^{3}{q}^{4}
\]
\[
-138{a}^{2}{b}^{5}m{q}^{6}-24{b}^{5}m{q}^{8}-216{a}^{5}
{b}^{4}{m}^{2}q+72{a}^{5}{b}^{4}{q}^{3}+40{a}^{3}{b}^{4}{m}^{4}q-
284{a}^{3}{b}^{4}{m}^{2}{q}^{3}+34{a}^{3}{b}^{4}{q}^{5}+104a{b}^
{4}{m}^{4}{q}^{3}
\]
\[
+16a{b}^{4}{m}^{2}{q}^{5}-32a{b}^{4}{q}^{7}+32{	a}^{6}{b}^{3}m-8{a}^{4}{b}^{3}{m}^{3}+52{a}^{4}{b}^{3}m{q}^{2}-108
{a}^{2}{b}^{3}{m}^{3}{q}^{2}+22{a}^{2}{b}^{3}m{q}^{4}+16{b}^{3}{m}^{3}{q}^{4}-8{b}^{3}m{q}^{6}
\]
\[
\left. +27{a}^{3}{b}^{2}{m}^{2}q-3{a}^{3}
{b}^{2}{q}^{3}-12a{b}^{2}{m}^{2}{q}^{3}-2{a}^{4}bm+6{a}^{2}bm{q}
^{2}-{a}^{3}q \right),
\]
\[
{\lambda}_2 = 6 \left( {m}^{2}-{a}^{2}-{q}^{2} \right)  \left( 32{a}^{5}{b}^{6}
{m}^{2}q+16{a}^{5}{b}^{6}{q}^{3}+8{a}^{3}{b}^{6}{m}^{4}q+44{a}^{
	3}{b}^{6}{m}^{2}{q}^{3}+16{a}^{3}{b}^{6}{q}^{5}-8{b}^{6}qa{m}^{6}-
4a{b}^{6}{m}^{4}{q}^{3} \right.
\]
\[
+14a{b}^{6}{m}^{2}{q}^{5}+3a{b}^{6}{q}^{7
}+48{a}^{4}{b}^{5}m{q}^{2}+48{a}^{2}{b}^{5}m{q}^{4}+48{a}^{5}{b}
^{4}q-12{a}^{3}{b}^{4}{m}^{2}q+48{a}^{3}{b}^{4}{q}^{3}+12a{b}^{4}{m}^{4}q+24a{b}^{4}{m}^{2}{q}^{3}
\]
\[
\left. +3a{b}^{4}{q}^{5}-16{a}^{4}{b}
^{3}m+16{a}^{2}{b}^{3}m{q}^{2}+8{b}^{3}m{q}^{4}-6a{b}^{2}{m}^{2}
q+9a{b}^{2}{q}^{3}+aq \right)  \left( 16{a}^{5}{b}^{4}-20{a}^{3}
{b}^{4}{m}^{2}+16{a}^{3}{b}^{4}{q}^{2} \right.
\]
\[
\left. +4a{b}^{4}{m}^{4}-12a{b}^{4}{m}^{2}{q}^{2}+3a{b}^{4}{q}^{4}+16{a}^{2}{b}^{3}mq-8{b}^{3}{m}
^{3}q+8m{b}^{3}{q}^{3}+6a{b}^{2}{q}^{2}+4mqb-a \right),
\]
\[
{\lambda}_1 = 4\sqrt {{a}^{2}-{m}^{2}+{q}^{2}} \left( 1024{a}^{11}{b}^{10}{m}^{2}q+256{a}^{11}{b}^{10}{q}^{3}-2240{a}^{9}{b}^{10}{m}^{4}q+2176{a}^{9}{b}^{10}{m}^{2}{q}^{3}+640{a}^{9}{b}^{10}{q}^{5} \right.
\]
\[
+2608{a}^{7}{b}^{10}{m}^{6}q-3824{a}^{7}{b}^{10}{m}^{4}{q}^{3}+1648{a}^{7}{b}^{10}{m}^{2}{q}^{5}+576{a}^{7}{b}^{10}{q}^{7}-1472{a}^{5}{b}^{10}{m}
^{8}q+3520{a}^{5}{b}^{10}{m}^{6}{q}^{3}
\]
\[
-2136{a}^{5}{b}^{10}{m}^{4}
{q}^{5}+544{a}^{5}{b}^{10}{m}^{2}{q}^{7}+232{a}^{5}{b}^{10}{q}^{9}
+368{a}^{3}{b}^{10}{m}^{10}q-1424{a}^{3}{b}^{10}{m}^{8}{q}^{3}+
1432{a}^{3}{b}^{10}{m}^{6}{q}^{5}-392{a}^{3}{b}^{10}{m}^{4}{q}^{7}
\]
\[
+79{a}^{3}{b}^{10}{m}^{2}{q}^{9}+43{a}^{3}{b}^{10}{q}^{11}-32a{b}^{10}{m}^{12}q+192a{b}^{10}{m}^{10}{q}^{3}-320a{b}^{10}{m}^{8}{q}
^{5}+160a{b}^{10}{m}^{6}{q}^{7}-2a{b}^{10}{m}^{4}{q}^{9}+4a{b}^{10}{m}^{2}{q}^{11}
\]
\[
+3a{b}^{10}{q}^{13}+1024{a}^{12}{b}^{9}m-1920{a}^{10}{b}^{9}{m}^{3}+3584{a}^{10}{b}^{9}m{q}^{2}+1184{a}^{8}{b}^{9}{m}^{5}-4160{a}^{8}{b}^{9}{m}^{3}{q}^{2}+4832{a}^{8}{b}^{9}m{q}^
{4}-320{a}^{6}{b}^{9}{m}^{7}
\]
\[
+1088{a}^{6}{b}^{9}{m}^{5}{q}^{2}-2832
{a}^{6}{b}^{9}{m}^{3}{q}^{4}+3056{a}^{6}{b}^{9}m{q}^{6}+32{a}^{4}{b}^{9}{m}^{9}+608{a}^{4}{b}^{9}{m}^{7}{q}^{2}-576{a}^{4}{b}^{9}{m}^{5}{q}^{4}-552{a}^{4}{b}^{9}{m}^{3}{q}^{6}
\]
\[
+890{a}^{4}{b}^{9}m{q}^{8}-416{a}^{2}{b}^{9}{m}^{9}{q}^{2}+944{a}^{2}{b}^{9}{m}^{7}{q}^{4}-640{a}^{2}{b}^{9}{m}^{5}{q}^{6}+28{a}^{2}{b}^{9}{m}^{3}{q}^{8}+110{a}^{2}{b}^{9}m{q}^{10}+64{b}^{9}{m}^{11}{q}^{2}
\]
\[
-224{b}^{9}{m}^{9}{q}^{4}+256{b}^{9}{m}^{7}{q}^{6}-88{b}^{9}{m}^{5}{q}^{8}-12
{b}^{9}{m}^{3}{q}^{10}+4{b}^{9}m{q}^{12}+768{a}^{11}{b}^{8}q+
1536{a}^{9}{b}^{8}{m}^{2}q+1920{a}^{9}{b}^{8}{q}^{3}
\]
\[
-2928{a}^{7}
{b}^{8}{m}^{4}q+4992{a}^{7}{b}^{8}{m}^{2}{q}^{3}+1728{a}^{7}{b}^{8}{q}^{5}+1872{a}^{5}{b}^{8}{m}^{6}q-6048{a}^{5}{b}^{8}{m}^{4}{q}^{3}+4920{a}^{5}{b}^{8}{m}^{2}{q}^{5}+696{a}^{5}{b}^{8}{q}^{7}
\]
\[
-528
{a}^{3}{b}^{8}{m}^{8}q+2640{a}^{3}{b}^{8}{m}^{6}{q}^{3}-3456{a}^{3
}{b}^{8}{m}^{4}{q}^{5}+1620{a}^{3}{b}^{8}{m}^{2}{q}^{7}+129{a}^{3}
{b}^{8}{q}^{9}+48a{b}^{8}{m}^{10}q-432a{b}^{8}{m}^{8}{q}^{3}
\]
\[
+744
a{b}^{8}{m}^{6}{q}^{5}-456a{b}^{8}{m}^{4}{q}^{7}+111a{b}^{8}{m}^{2
}{q}^{9}+9a{b}^{8}{q}^{11}+256{a}^{10}{b}^{7}m-640{a}^{8}{b}^{7}
{m}^{3}+2816{a}^{8}{b}^{7}m{q}^{2}+288{a}^{6}{b}^{7}{m}^{5}
\]
\[
-1552
{a}^{6}{b}^{7}{m}^{3}{q}^{2}+4512{a}^{6}{b}^{7}m{q}^{4}-32{a}^{4}{b}^{7}{m}^{7}-672{a}^{4}{b}^{7}{m}^{5}{q}^{2}-576{a}^{4}{b}^{7}{m}^{3}{q}^{4}+2252{a}^{4}{b}^{7}m{q}^{6}+656{a}^{2}{b}^{7}{m}^{7}{q}
^{2}
\]
\[
-1128{a}^{2}{b}^{7}{m}^{5}{q}^{4}+308{a}^{2}{b}^{7}{m}^{3}{q}^
{6}+302{a}^{2}{b}^{7}m{q}^{8}-128{b}^{7}{m}^{9}{q}^{2}+352{b}^{7}{m}^{7}{q}^{4}-312{b}^{7}{m}^{5}{q}^{6}+80{b}^{7}{m}^{3}{q}^{8}+10{b}^{7}m{q}^{10}
\]
\[
+256{a}^{9}{b}^{6}q+464{a}^{7}{b}^{6}{m}^{2}q+
896{a}^{7}{b}^{6}{q}^{3}-648{a}^{5}{b}^{6}{m}^{4}q+2336{a}^{5}{b}^{6}{m}^{2}{q}^{3}+792{a}^{5}{b}^{6}{q}^{5}+232{a}^{3}{b}^{6}{m}^{6}q-1752{a}^{3}{b}^{6}{m}^{4}{q}^{3}
\]
\[
+1954{a}^{3}{b}^{6}{m}^{2}{q}
^{5}+162{a}^{3}{b}^{6}{q}^{7}-16a{b}^{6}{m}^{8}q+352a{b}^{6}{m}^
{6}{q}^{3}-660a{b}^{6}{m}^{4}{q}^{5}+368a{b}^{6}{m}^{2}{q}^{7}+10
a{b}^{6}{q}^{9}-96{a}^{8}{b}^{5}m
\]
\[
+336{a}^{6}{b}^{5}m{q}^{2}+240
{a}^{4}{b}^{5}{m}^{3}{q}^{2}+984{a}^{4}{b}^{5}m{q}^{4}-360{a}^{2}{b}^{5}{m}^{5}{q}^{2}+228{a}^{2}{b}^{5}{m}^{3}{q}^{4}+342{a}^{2}{b}^{5}m{q}^{6}+96{b}^{5}{m}^{7}{q}^{2}-192{b}^{5}{m}^{5}{q}^{4}
\]
\[
+96
{b}^{5}{m}^{3}{q}^{6}+6{b}^{5}m{q}^{8}+40{a}^{5}{b}^{4}{m}^{2}q+
72{a}^{5}{b}^{4}{q}^{3}-16{a}^{3}{b}^{4}{m}^{4}q+388{a}^{3}{b}^{
	4}{m}^{2}{q}^{3}+110{a}^{3}{b}^{4}{q}^{5}-8a{b}^{4}{m}^{6}q-120a
{b}^{4}{m}^{4}{q}^{3}
\]
\[
+178a{b}^{4}{m}^{2}{q}^{5}+6a{b}^{4}{q}^{7}-
32{a}^{6}{b}^{3}m+8{a}^{4}{b}^{3}{m}^{3}-28{a}^{4}{b}^{3}m{q}^{2
}+76{a}^{2}{b}^{3}{m}^{3}{q}^{2}+50{a}^{2}{b}^{3}m{q}^{4}-32{b}^
{3}{m}^{5}{q}^{2}
\]
\[ +40{b}^{3}{m}^{3}{q}^{4}-2{b}^{3}m{q}^{6}-9{a}^
{3}{b}^{2}{m}^{2}q+3{a}^{3}{b}^{2}{q}^{3}+6a{b}^{2}{m}^{4}q+12a{
	b}^{2}{m}^{2}{q}^{3}+3a{b}^{2}{q}^{5}-2{a}^{4}bm
\]
\[\left. 
-4{a}^{2}bm{q}^{
	2}+4b{m}^{3}{q}^{2}-2bm{q}^{4}+{a}^{3}q-a{m}^{2}q+a{q}^{3}
\right),
\]
\[
{\lambda}_0 = - \left( 96{a}^{6}{b}^{6}{m}^{2}q+16{a}^{6}{b}^{6}{q}^{3}-176{a}
^{4}{b}^{6}{m}^{4}q+100{a}^{4}{b}^{6}{m}^{2}{q}^{3}+24{a}^{4}{b}^{
	6}{q}^{5}+96{a}^{2}{b}^{6}{m}^{6}q-140{a}^{2}{b}^{6}{m}^{4}{q}^{3} \right.
\]
\[
+20{a}^{2}{b}^{6}{m}^{2}{q}^{5}+9{a}^{2}{b}^{6}{q}^{7}-16{b}^{6}
{m}^{8}q+40{b}^{6}{m}^{6}{q}^{3}-24{b}^{6}{m}^{4}{q}^{5}-2{b}^{6
}{m}^{2}{q}^{7}+{b}^{6}{q}^{9}+128{a}^{7}{b}^{5}m-80{a}^{5}{b}^{5}
{m}^{3}
\]
\[
+272{a}^{5}{b}^{5}m{q}^{2}+16{a}^{3}{b}^{5}{m}^{5}-160{a}
^{3}{b}^{5}{m}^{3}{q}^{2}+148m{a}^{3}{b}^{5}{q}^{4}+32a{b}^{5}{m}^
{5}{q}^{2}-48a{b}^{5}{m}^{3}{q}^{4}+16ma{b}^{5}{q}^{6}+48{a}^{6}
{b}^{4}q
\]
\[
+108{a}^{4}{b}^{4}{m}^{2}q+64{a}^{4}{b}^{4}{q}^{3}-100{a}^{2}{b}^{4}{m}^{4}q+96{a}^{2}{b}^{4}{m}^{2}{q}^{3}+13{a}^{2}{b}^{4}{q}^{5}+24{b}^{4}{m}^{6}q-44{b}^{4}{m}^{4}{q}^{3}+18{b}^{4}{m}
^{2}{q}^{5}+{b}^{4}{q}^{7}
\]
\[
+48m{a}^{5}{b}^{3}-16{a}^{3}{b}^{3}{m}^{
	3}+120{a}^{3}{b}^{3}m{q}^{2}-32a{b}^{3}{m}^{3}{q}^{2}+40ma{b}^{3
}{q}^{4}+8{a}^{4}{b}^{2}q+28{a}^{2}{b}^{2}{m}^{2}q+11{a}^{2}{b}^
{2}{q}^{3}-12{b}^{2}{m}^{4}q
\]
\[\left.
+14{b}^{2}{m}^{2}{q}^{3}-{b}^{2}{q}^{5
}+4{a}^{3}bm+8abm{q}^{2}-{a}^{2}q+2{m}^{2}q-{q}^{3} \right) 
\left( 16{a}^{6}{b}^{4}+4{a}^{4}{b}^{4}{m}^{2}+24{a}^{4}{b}^{4}
{q}^{2}-4{a}^{2}{b}^{4}{m}^{4} \right.
\]
\[
+8{a}^{2}{b}^{4}{m}^{2}{q}^{2}+9{a
}^{2}{b}^{4}{q}^{4}-4{b}^{4}{m}^{4}{q}^{2}+4{b}^{4}{m}^{2}{q}^{4}+
{b}^{4}{q}^{6}+24{a}^{3}{b}^{3}mq-8a{b}^{3}{m}^{3}q+12a{b}^{3}m{q}^{3}+8{a}^{4}{b}^{2}
\]
\[\left.
+14{a}^{2}{b}^{2}{q}^{2}+2{b}^{2}{q}^{4}+4
abmq+{a}^{2}+{q}^{2} \right),
\]
\[
{\tilde {\lambda}}_4 =\left( 16{a}^{6}{b}^{4}+4{a}^{4}{b}^{4}{m}^{2}+24{a}^{4}{b}^{4}{q}^{2}-4{a}^{2}{b}^{4}{m}^{4}+4{a}^{2}{b}^{4}{m}^{2}{q}^{2}+9{a}^{2}{b}^{4}{q}^{4}+8{a}^{3}{b}^{3}mq+8a{b}^{3}{m}^{3}q \right.
\]
\[
\left. +4a{b}^{3}m{q}^{3}-8{a}^{4}{b}^{2}-2{a}^{2}{b}^{2}{q}^{2}+4{b}^{2}{q}^{4}
-4abmq+{a}^{2} \right) ^{2},
\]
\[
{\tilde {\lambda}}_3 = -4\sqrt {{a}^{2}-{m}^{2}+{q}^{2}} \left( 16{a}^{5}{b}^{4}-20{a}^
{3}{b}^{4}{m}^{2}+16{a}^{3}{b}^{4}{q}^{2}+4a{b}^{4}{m}^{4}-12a{b
}^{4}{m}^{2}{q}^{2}+3a{b}^{4}{q}^{4} \right.
\]
\[
\left. +16{a}^{2}{b}^{3}mq-8{b}^{3}
{m}^{3}q+8m{b}^{3}{q}^{3}+6a{b}^{2}{q}^{2}+4mqb-a \right) 
\left( 16{a}^{6}{b}^{4}+4{a}^{4}{b}^{4}{m}^{2} +24{a}^{4}{b}^{4}
{q}^{2} \right. 
\]
\[
-4{a}^{2}{b}^{4}{m}^{4}+4{a}^{2}{b}^{4}{m}^{2}{q}^{2}+9{a}^{2}{b}^{4}{q}^{4}+8{a}^{3}{b}^{3}mq+8a{b}^{3}{m}^{3}q+4a{b}^{3}m{q}^{3}-8{a}^{4}{b}^{2}
\]
\[\left. 
-2{a}^{2}{b}^{2}{q}^{2}+4{b}^{2}{q}^{4}
-4abmq+{a}^{2} \right), 
\]
\[
{\tilde {\lambda}}_2 = 1536{a}^{12}{b}^{8}-3328{a}^{10}{b}^{8}{m}^{2}+4608{a}^{10}{b}^{8}{q}^{2}+4448{a}^{8}{b}^{8}{m}^{4}-7936{a}^{8}{b}^{8}{m}^{2}{q}^{2}+5184{a}^{8}{b}^{8}{q}^{4}
\]
\[
-2816{a}^{6}{b}^{8}{m}^{6}+8224{a}^{
	6}{b}^{8}{m}^{4}{q}^{2}-6672{a}^{6}{b}^{8}{m}^{2}{q}^{4}+2688{a}^{
	6}{b}^{8}{q}^{6}+736{a}^{4}{b}^{8}{m}^{8}-3584{a}^{4}{b}^{8}{m}^{6
}{q}^{2}+4880{a}^{4}{b}^{8}{m}^{4}{q}^{4}
\]
\[
-2272{a}^{4}{b}^{8}{m}^{2
}{q}^{6}+630{a}^{4}{b}^{8}{q}^{8}-64{a}^{2}{b}^{8}{m}^{10}+480{a}^{2}{b}^{8}{m}^{8}{q}^{2}-1120{a}^{2}{b}^{8}{m}^{6}{q}^{4}+912{a}
^{2}{b}^{8}{m}^{4}{q}^{6}-244{a}^{2}{b}^{8}{m}^{2}{q}^{8}
\]
\[
+54{a}^{2}{b}^{8}{q}^{10}+3072{a}^{9}{b}^{7}mq-5376{a}^{7}{b}^{7}{m}^{3}q+
7168{a}^{7}{b}^{7}m{q}^{3}+5120{a}^{5}{b}^{7}{m}^{5}q-10048{a}^{5}{b}^{7}{m}^{3}{q}^{3}+5824{a}^{5}{b}^{7}m{q}^{5}
\]
\[
-2048{a}^{3}{b}^
{7}{m}^{7}q+6528{a}^{3}{b}^{7}{m}^{5}{q}^{3}-5984{a}^{3}{b}^{7}{m}
^{3}{q}^{5}+1904{a}^{3}{b}^{7}m{q}^{7}+256a{b}^{7}{m}^{9}q-1344a
{b}^{7}{m}^{7}{q}^{3}+2016a{b}^{7}{m}^{5}{q}^{5}
\]
\[
-1104a{b}^{7}{m}^{3}{q}^{7}+200a{b}^{7}m{q}^{9}+1152{a}^{8}{b}^{6}{q}^{2}-288{a}^{6}{b}^{6}{m}^{2}{q}^{2}+2304{a}^{6}{b}^{6}{q}^{4}-672{a}^{4}{b}^{6}{m}^{4}{q}^{2}+192{a}^{4}{b}^{6}{m}^{2}{q}^{4}
\]
\[
+1400{a}^{4}{b}^{6}
{q}^{6}+960{a}^{2}{b}^{6}{m}^{6}{q}^{2}-1696{a}^{2}{b}^{6}{m}^{4}{q}^{4}+720{a}^{2}{b}^{6}{m}^{2}{q}^{6}+248{a}^{2}{b}^{6}{q}^{8}-256{b}^{6}{m}^{8}{q}^{2}+768{b}^{6}{m}^{6}{q}^{4}
\]
\[
-800{b}^{6}{m}^
{4}{q}^{6}+288{b}^{6}{m}^{2}{q}^{8}+8{b}^{6}{q}^{10}+256{a}^{7}{b}^{5}mq-896{a}^{5}{b}^{5}{m}^{3}q+1792{a}^{5}{b}^{5}m{q}^{3}+768
{a}^{3}{b}^{5}{m}^{5}q
\]
\[
-2464{a}^{3}{b}^{5}{m}^{3}{q}^{3}+2048{a}^
{3}{b}^{5}m{q}^{5}-128a{b}^{5}{m}^{7}q+928a{b}^{5}{m}^{5}{q}^{3}-
1312a{b}^{5}{m}^{3}{q}^{5}+584a{b}^{5}m{q}^{7}-192{a}^{8}{b}^{4}
\]
\[
+304{a}^{6}{b}^{4}{m}^{2}-384{a}^{6}{b}^{4}{q}^{2}-208{a}^{4}{b}
^{4}{m}^{4}+800{a}^{4}{b}^{4}{m}^{2}{q}^{2}+52{a}^{4}{b}^{4}{q}^{4
}+32{a}^{2}{b}^{4}{m}^{6}-784{a}^{2}{b}^{4}{m}^{4}{q}^{2}
\]
\[
+696{a}
^{2}{b}^{4}{m}^{2}{q}^{4}+244{a}^{2}{b}^{4}{q}^{6}+256{b}^{4}{m}^{6}{q}^{2}-512{b}^{4}{m}^{4}{q}^{4}+256{b}^{4}{m}^{2}{q}^{6}+16{b}^{4}{q}^{8}-192{a}^{5}{b}^{3}mq+192{a}^{3}{b}^{3}{m}^{3}q
\]
\[
-80{a}
^{3}{b}^{3}m{q}^{3}-64a{b}^{3}{m}^{5}q-48a{b}^{3}{m}^{3}{q}^{3}+
152a{b}^{3}m{q}^{5}-40{a}^{4}{b}^{2}{q}^{2}+80{a}^{2}{b}^{2}{m}^
{2}{q}^{2}-40{a}^{2}{b}^{2}{q}^{4}
\]
\[
-64{b}^{2}{m}^{4}{q}^{2}+64{b}
^{2}{m}^{2}{q}^{4}+8{b}^{2}{q}^{6}-32{a}^{3}bmq+32ab{m}^{3}q-40
abm{q}^{3}+6{a}^{4}-4{a}^{2}{m}^{2}+6{a}^{2}{q}^{2},
\]
\[
{\tilde {\lambda}}_1 = -4 \sqrt {{a}^{2}-{m}^{2}+{q}^{2}} \left( 16{a}^{6}{b}^{4}+4{a}^{4}{b}^{4}{m}^{2}+24{a}^{4}{b}
^{4}{q}^{2}-4{a}^{2}{b}^{4}{m}^{4}+8{a}^{2}{b}^{4}{m}^{2}{q}^{2}+9
{a}^{2}{b}^{4}{q}^{4} -4{b}^{4}{m}^{4}{q}^{2}\right.
\]
\[\left. 
+4{b}^{4}{m}^{2}{q}^
{4}+{b}^{4}{q}^{6}+24{a}^{3}{b}^{3}mq-8a{b}^{3}{m}^{3}q+12a{b}^{3}m{q}^{3}+8{a}^{4}{b}^{2}+14{a}^{2}{b}^{2}{q}^{2}+2{b}^{2}{q}^{
	4}+4abmq+{a}^{2}+{q}^{2} \right)
\]
\[
\times  \left( 16{a}^{5}{b}^{4}-20{a}^{3}{b}^{4}{m}^{2}+16{a}^{3}{b}^{4
}{q}^{2}+4a{b}^{4}{m}^{4}-12a{b}^{4}{m}^{2}{q}^{2}+3a{b}^{4}{q}^
{4}+16{a}^{2}{b}^{3}mq \right.
\]
\[
\left. -8{b}^{3}{m}^{3}q+8m{b}^{3}{q}^{3}+6a{b}
^{2}{q}^{2}+4mqb-a \right),
\]
\[
{\tilde {\lambda}}_0 = \left( 16{a}^{6}{b}^{4}+4{a}^{4}{b}^{4}{m}^{2}+24{a}^{4}{b}^{4}{q}^{2}-4{a}^{2}{b}^{4}{m}^{4}+8{a}^{2}{b}^{4}{m}^{2}{q}^{2}+9{a
}^{2}{b}^{4}{q}^{4}-4{b}^{4}{m}^{4}{q}^{2}+4{b}^{4}{m}^{2}{q}^{4}+
{b}^{4}{q}^{6} \right.
\]
\[
\left. +24{a}^{3}{b}^{3}mq-8a{b}^{3}{m}^{3}q+12a{b}^{3}m{
	q}^{3}+8{a}^{4}{b}^{2}+14{a}^{2}{b}^{2}{q}^{2}+2{b}^{2}{q}^{4}+4
abmq+{a}^{2}+{q}^{2} \right) ^{2}.
\]
\be
\ee

\section{Functions in stretched horizon method}

\begin{eqnarray}
g_1^2&=&\frac{1}{{a}^{2}{x}^{2
	}+2\,alx+{l}^{2}+{r}^{2}}
\left( {m}^{2}+ \left( ax+l \right) ^{2} \right) 
\{ {a}^{6}{b}^{4}{x}^{6}-2\,{a}^{4}{b}^{4}{l}^{2}{x}^{6}-4\,{a}^{
	4}{b}^{4}mr{x}^{6}+2\,{a}^{4}{b}^{4}{q}^{2}{x}^{6} \nn\\
&+& 2\,{a}^{4}{b}^{4}{r
}^{2}{x}^{6}+{a}^{2}{b}^{4}{l}^{4}{x}^{6}+4\,{a}^{2}{b}^{4}{l}^{2}mr{x
}^{6}-2\,{a}^{2}{b}^{4}{l}^{2}{q}^{2}{x}^{6}-2\,{a}^{2}{b}^{4}{l}^{2}{
	r}^{2}{x}^{6}+4\,{a}^{2}{b}^{4}{m}^{2}{r}^{2}{x}^{6}-4\,{a}^{2}{b}^{4}
m{q}^{2}r{x}^{6}\nn\\
&-&4\,{a}^{2}{b}^{4}m{r}^{3}{x}^{6}+{a}^{2}{b}^{4}{q}^{4
}{x}^{6}+2\,{a}^{2}{b}^{4}{q}^{2}{r}^{2}{x}^{6}+{a}^{2}{b}^{4}{r}^{4}{
	x}^{6}+6\,{a}^{5}{b}^{4}l{x}^{5}-12\,{a}^{3}{b}^{4}{l}^{3}{x}^{5}-24\,
{a}^{3}{b}^{4}lmr{x}^{5}\nn\\
&+&12\,{a}^{3}{b}^{4}l{q}^{2}{x}^{5}+12\,{a}^{3}
{b}^{4}l{r}^{2}{x}^{5}+6\,a{b}^{4}{l}^{5}{x}^{5}+24\,a{b}^{4}{l}^{3}mr
{x}^{5}-12\,a{b}^{4}{l}^{3}{q}^{2}{x}^{5}-12\,a{b}^{4}{l}^{3}{r}^{2}{x
}^{5}\nn\\
&+&24\,a{b}^{4}l{m}^{2}{r}^{2}{x}^{5}-24\,a{b}^{4}lm{q}^{2}r{x}^{5}
-24\,a{b}^{4}lm{r}^{3}{x}^{5}+6\,a{b}^{4}l{q}^{4}{x}^{5}+12\,a{b}^{4}l
{q}^{2}{r}^{2}{x}^{5}+6\,a{b}^{4}l{r}^{4}{x}^{5}-2\,{a}^{6}{b}^{4}{x}^
{4}\nn\\
&+&21\,{a}^{4}{b}^{4}{l}^{2}{x}^{4}+4\,{a}^{4}{b}^{4}{m}^{2}{x}^{4}+
12\,{a}^{4}{b}^{4}mr{x}^{4}-6\,{a}^{4}{b}^{4}{q}^{2}{x}^{4}-3\,{a}^{4}
{b}^{4}{r}^{2}{x}^{4}-28\,{a}^{2}{b}^{4}{l}^{4}{x}^{4}+8\,{a}^{2}{b}^{
	4}{l}^{2}{m}^{2}{x}^{4}\nn\\
&-&84\,{a}^{2}{b}^{4}{l}^{2}mr{x}^{4}+32\,{a}^{2}
{b}^{4}{l}^{2}{q}^{2}{x}^{4}+36\,{a}^{2}{b}^{4}{l}^{2}{r}^{2}{x}^{4}-
24\,{a}^{2}{b}^{4}{m}^{2}{r}^{2}{x}^{4}+24\,{a}^{2}{b}^{4}m{q}^{2}r{x}
^{4}+12\,{a}^{2}{b}^{4}m{r}^{3}{x}^{4}\nn\\
&-&4\,{a}^{2}{b}^{4}{q}^{4}{x}^{4}
-8\,{a}^{2}{b}^{4}{q}^{2}{r}^{2}{x}^{4}+9\,{b}^{4}{l}^{6}{x}^{4}+4\,{b
}^{4}{l}^{4}{m}^{2}{x}^{4}+24\,{b}^{4}{l}^{4}mr{x}^{4}-18\,{b}^{4}{l}^
{4}{q}^{2}{x}^{4}-9\,{b}^{4}{l}^{4}{r}^{2}{x}^{4}\nn\\
&+&36\,{b}^{4}{l}^{2}{m
}^{2}{r}^{2}{x}^{4}-32\,{b}^{4}{l}^{2}m{q}^{2}r{x}^{4}-40\,{b}^{4}{l}^
{2}m{r}^{3}{x}^{4}+9\,{b}^{4}{l}^{2}{q}^{4}{x}^{4}+12\,{b}^{4}{l}^{2}{
	q}^{2}{r}^{2}{x}^{4}+15\,{b}^{4}{l}^{2}{r}^{4}{x}^{4}+{b}^{4}{q}^{4}{r
}^{2}{x}^{4}\nn\\
&-&2\,{b}^{4}{q}^{2}{r}^{4}{x}^{4}+{b}^{4}{r}^{6}{x}^{4}-12
\,{a}^{5}{b}^{4}l{x}^{3}+40\,{a}^{3}{b}^{4}{l}^{3}{x}^{3}-8\,{a}^{3}{b
}^{4}l{m}^{2}{x}^{3}+80\,{a}^{3}{b}^{4}lmr{x}^{3}-24\,{a}^{3}{b}^{4}l{
	q}^{2}{x}^{3}\nn\\
&-&24\,{a}^{3}{b}^{4}l{r}^{2}{x}^{3}-28\,a{b}^{4}{l}^{5}{x}
^{3}-8\,a{b}^{4}{l}^{3}{m}^{2}{x}^{3}-64\,a{b}^{4}{l}^{3}mr{x}^{3}+40
\,a{b}^{4}{l}^{3}{q}^{2}{x}^{3}+8\,a{b}^{4}{l}^{3}{r}^{2}{x}^{3}-72\,a
{b}^{4}l{m}^{2}{r}^{2}{x}^{3}\nn\\
&+&56\,a{b}^{4}lm{q}^{2}r{x}^{3}+48\,a{b}^{
	4}lm{r}^{3}{x}^{3}-12\,a{b}^{4}l{q}^{4}{x}^{3}-8\,a{b}^{4}l{q}^{2}{r}^
{2}{x}^{3}-12\,a{b}^{4}l{r}^{4}{x}^{3}+{a}^{6}{b}^{4}{x}^{2}-28\,{a}^{
	4}{b}^{4}{l}^{2}{x}^{2}\nn\\
&+&8\,{a}^{4}{b}^{4}{m}^{2}{x}^{2}-12\,{a}^{4}{b}
^{4}mr{x}^{2}+4\,{a}^{4}{b}^{4}{q}^{2}{x}^{2}+4\,{a}^{3}{b}^{3}qr{x}^{
	4}+37\,{a}^{2}{b}^{4}{l}^{4}{x}^{2}+12\,{a}^{2}{b}^{4}{l}^{2}{m}^{2}{x
}^{2}+84\,{a}^{2}{b}^{4}{l}^{2}mr{x}^{2}\nn\\
&-&38\,{a}^{2}{b}^{4}{l}^{2}{q}^
{2}{x}^{2}-18\,{a}^{2}{b}^{4}{l}^{2}{r}^{2}{x}^{2}+36\,{a}^{2}{b}^{4}{
	m}^{2}{r}^{2}{x}^{2}-20\,{a}^{2}{b}^{4}m{q}^{2}r{x}^{2}-12\,{a}^{2}{b}
^{4}m{r}^{3}{x}^{2}+4\,{a}^{2}{b}^{4}{q}^{4}{x}^{2}\nn\\
&+&6\,{a}^{2}{b}^{4}{
	q}^{2}{r}^{2}{x}^{2}-3\,{a}^{2}{b}^{4}{r}^{4}{x}^{2}-4\,a{b}^{3}{l}^{2
}qr{x}^{4}-8\,a{b}^{3}mq{r}^{2}{x}^{4}+4\,a{b}^{3}{q}^{3}r{x}^{4}+4\,a
{b}^{3}q{r}^{3}{x}^{4}+6\,{b}^{4}{l}^{6}{x}^{2}\nn\\
&+&24\,{b}^{4}{l}^{4}mr{x
}^{2}-6\,{b}^{4}{l}^{4}{q}^{2}{x}^{2}-30\,{b}^{4}{l}^{4}{r}^{2}{x}^{2}
-8\,{b}^{4}{l}^{2}m{r}^{3}{x}^{2}+12\,{b}^{4}{l}^{2}{q}^{2}{r}^{2}{x}^
{2}-6\,{b}^{4}{l}^{2}{r}^{4}{x}^{2}+2\,{b}^{4}{q}^{2}{r}^{4}{x}^{2}\nn\\
&-&2
\,{b}^{4}{r}^{6}{x}^{2}+6\,{a}^{5}{b}^{4}lx-28\,{a}^{3}{b}^{4}{l}^{3}x
-8\,{a}^{3}{b}^{4}l{m}^{2}x-24\,{a}^{3}{b}^{4}lmrx+12\,{a}^{3}{b}^{4}l
{q}^{2}x+12\,{a}^{3}{b}^{4}l{r}^{2}x\nn\\
&-&8\,{a}^{2}{b}^{3}lmq{x}^{3}+16\,{
	a}^{2}{b}^{3}lqr{x}^{3}-10\,a{b}^{4}{l}^{5}x-24\,a{b}^{4}{l}^{3}mrx+4
\,a{b}^{4}{l}^{3}{q}^{2}x+36\,a{b}^{4}{l}^{3}{r}^{2}x+8\,a{b}^{4}lm{r}
^{3}x\nn\\
&-&4\,a{b}^{4}l{q}^{2}{r}^{2}x+6\,a{b}^{4}l{r}^{4}x-8\,{b}^{3}{l}^{
	3}mq{x}^{3}-24\,{b}^{3}lmq{r}^{2}{x}^{3}+8\,{b}^{3}l{q}^{3}r{x}^{3}+16
\,{b}^{3}lq{r}^{3}{x}^{3}+9\,{a}^{4}{b}^{4}{l}^{2}+4\,{a}^{4}{b}^{4}{m
}^{2}\nn\\
&+&4\,{a}^{4}{b}^{4}mr+{a}^{4}{b}^{4}{r}^{2}-2\,{a}^{4}{b}^{2}{x}^{
	4}+8\,{a}^{3}{b}^{3}mq{x}^{2}-8\,{a}^{3}{b}^{3}qr{x}^{2}+6\,{a}^{2}{b}
^{4}{l}^{4}+12\,{a}^{2}{b}^{4}{l}^{2}mr+4\,{a}^{2}{b}^{4}m{r}^{3}\nn\\
&+&2\,{
	a}^{2}{b}^{4}{r}^{4}+2\,{a}^{2}{b}^{2}{l}^{2}{x}^{4}+4\,{a}^{2}{b}^{2}
mr{x}^{4}-2\,{a}^{2}{b}^{2}{q}^{2}{x}^{4}-2\,{a}^{2}{b}^{2}{r}^{2}{x}^
{4}+16\,a{b}^{3}{l}^{2}mq{x}^{2}-8\,a{b}^{3}{l}^{2}qr{x}^{2}\nn\\
&+&24\,a{b}^
{3}mq{r}^{2}{x}^{2}-4\,a{b}^{3}{q}^{3}r{x}^{2}-8\,a{b}^{3}q{r}^{3}{x}^
{2}+{b}^{4}{l}^{6}+7\,{b}^{4}{l}^{4}{r}^{2}+7\,{b}^{4}{l}^{2}{r}^{4}+{
	b}^{4}{r}^{6}-8\,{a}^{3}{b}^{2}l{x}^{3}-16\,{a}^{2}{b}^{3}lmqx\nn\\
&+&8\,a{b}
^{2}{l}^{3}{x}^{3}+16\,a{b}^{2}lmr{x}^{3}-8\,a{b}^{2}l{q}^{2}{x}^{3}-8
\,a{b}^{2}l{r}^{2}{x}^{3}-16\,{b}^{3}{l}^{3}qrx+2\,{a}^{4}{b}^{2}{x}^{
	2}+8\,{a}^{3}{b}^{3}mq+4\,{a}^{3}{b}^{3}qr\nn\\
&-&16\,{a}^{2}{b}^{2}{l}^{2}{x
}^{2}-8\,{a}^{2}{b}^{2}mr{x}^{2}+4\,{a}^{2}{b}^{2}{q}^{2}{x}^{2}+12\,a
{b}^{3}{l}^{2}qr+4\,a{b}^{3}q{r}^{3}+6\,{b}^{2}{l}^{4}{x}^{2}+16\,{b}^
{2}{l}^{2}mr{x}^{2}-2\,{b}^{2}{l}^{2}{q}^{2}{x}^{2}\nn\\
&-&12\,{b}^{2}{l}^{2}
{r}^{2}{x}^{2}+6\,{b}^{2}{q}^{2}{r}^{2}{x}^{2}-2\,{b}^{2}{r}^{4}{x}^{2
}+8\,{a}^{3}{b}^{2}lx-8\,a{b}^{2}{l}^{3}x-16\,a{b}^{2}lmrx-4\,a{b}^{2}
l{q}^{2}x+8\,a{b}^{2}l{r}^{2}x\nn\\
&+&6\,{a}^{2}{b}^{2}{l}^{2}+4\,{a}^{2}{b}^
{2}mr+4\,{a}^{2}{b}^{2}{q}^{2}+2\,{a}^{2}{b}^{2}{r}^{2}-4\,abqr{x}^{2}
+2\,{b}^{2}{l}^{4}+4\,{b}^{2}{l}^{2}{r}^{2}+2\,{b}^{2}{r}^{4}-8\,blqrx
+{a}^{2}{x}^{2}\nn\\
&+&4\,abqr+2\,alx+{l}^{2}+{r}^{2} \}. \label{g1}
\end{eqnarray}

\begin{eqnarray}
g_2&=&\frac{32}{4 \left( {a}^{2}+\frac{1}{2}\,{q}^{2}
	\right) ^{2}}\,\{ -\,a \left( -\frac{1}{16}+ \left( \frac{3}{16}\,{q}^{4}+ \left( {a}^{2}
+\frac{1}{4}\,{l}^{2} \right) {q}^{2}+{a}^{4}+\frac{3}{4}\,{a}^{2}{l}^{2}+\frac{1}{4}\,{l}^{4}
\right) {b}^{4}+\frac{3}{8}\,{b}^{2}{q}^{2} \right) \nn\\
&\times&
\sqrt {{a}^{2}-{l}^{2}+{q
	}^{2}}-\,qb \left(  \left( \frac{1}{8}\,{q}^{4}+ \left( {a}^{2}+\frac{1}{4}\,{l}^{2}
\right) {q}^{2}+{a}^{4}-\frac{1}{2}\,{a}^{2}{l}^{2}-\frac{1}{2}\,{l}^{4} \right) {b}^
{2}-\frac{1}{4}\,{l}^{2}+\frac{1}{8}\,{q}^{2} \right) \}.\nn\\
&&\label{g2}
\end{eqnarray}

\begin{eqnarray}
g_3^2&=&  -Z^{-1}\Delta_x \{ {a}^{6}
{b}^{4}{x}^{6}-2{a}^{4}{b}^{4}{l}^{2}{x}^{6}-4{a}^{4}{b}^{4}mr{x}^
{6}+2{a}^{4}{b}^{4}{q}^{2}{x}^{6}+2{a}^{4}{b}^{4}{r}^{2}{x}^{6}+{a
}^{2}{b}^{4}{l}^{4}{x}^{6}\nn\\
&+&4{a}^{2}{b}^{4}{l}^{2}mr{x}^{6}-2{a}^{2
}{b}^{4}{l}^{2}{q}^{2}{x}^{6}-2{a}^{2}{b}^{4}{l}^{2}{r}^{2}{x}^{6}+4
{a}^{2}{b}^{4}{m}^{2}{r}^{2}{x}^{6}-4{a}^{2}{b}^{4}m{q}^{2}r{x}^{6
}-4{a}^{2}{b}^{4}m{r}^{3}{x}^{6}\nn\\
&+&{a}^{2}{b}^{4}{q}^{4}{x}^{6}+2{a}
^{2}{b}^{4}{q}^{2}{r}^{2}{x}^{6}+{a}^{2}{b}^{4}{r}^{4}{x}^{6}+6{a}^{
	5}{b}^{4}l{x}^{5}-12{a}^{3}{b}^{4}{l}^{3}{x}^{5}-24{a}^{3}{b}^{4}l
mr{x}^{5}+12{a}^{3}{b}^{4}l{q}^{2}{x}^{5}\nn\\
&+&12{a}^{3}{b}^{4}l{r}^{2}
{x}^{5}+6a{b}^{4}{l}^{5}{x}^{5}+24a{b}^{4}{l}^{3}mr{x}^{5}-12a{b
}^{4}{l}^{3}{q}^{2}{x}^{5}-12a{b}^{4}{l}^{3}{r}^{2}{x}^{5}+24a{b}^
{4}l{m}^{2}{r}^{2}{x}^{5}\nn\\
&-&24a{b}^{4}lm{q}^{2}r{x}^{5}-24a{b}^{4}lm
{r}^{3}{x}^{5}+6a{b}^{4}l{q}^{4}{x}^{5}+12a{b}^{4}l{q}^{2}{r}^{2}{
	x}^{5}+6a{b}^{4}l{r}^{4}{x}^{5}-2{a}^{6}{b}^{4}{x}^{4}+21{a}^{4}
{b}^{4}{l}^{2}{x}^{4}\nn\\
&+&4{a}^{4}{b}^{4}{m}^{2}{x}^{4}+12{a}^{4}{b}^{
	4}mr{x}^{4}-6{a}^{4}{b}^{4}{q}^{2}{x}^{4}-3{a}^{4}{b}^{4}{r}^{2}{x
}^{4}-28{a}^{2}{b}^{4}{l}^{4}{x}^{4}+8{a}^{2}{b}^{4}{l}^{2}{m}^{2}
{x}^{4}-84{a}^{2}{b}^{4}{l}^{2}mr{x}^{4}\nn\\
&+&32{a}^{2}{b}^{4}{l}^{2}{q
}^{2}{x}^{4}+36{a}^{2}{b}^{4}{l}^{2}{r}^{2}{x}^{4}-24{a}^{2}{b}^{4
}{m}^{2}{r}^{2}{x}^{4}+24{a}^{2}{b}^{4}m{q}^{2}r{x}^{4}+12{a}^{2}{
	b}^{4}m{r}^{3}{x}^{4}-4{a}^{2}{b}^{4}{q}^{4}{x}^{4}\nn\\
&-&8{a}^{2}{b}^{4
}{q}^{2}{r}^{2}{x}^{4}+9{b}^{4}{l}^{6}{x}^{4}+4{b}^{4}{l}^{4}{m}^{
	2}{x}^{4}+24{b}^{4}{l}^{4}mr{x}^{4}-18{b}^{4}{l}^{4}{q}^{2}{x}^{4}
-9{b}^{4}{l}^{4}{r}^{2}{x}^{4}+36{b}^{4}{l}^{2}{m}^{2}{r}^{2}{x}^{
	4}\nn\\
&-&32{b}^{4}{l}^{2}m{q}^{2}r{x}^{4}-40{b}^{4}{l}^{2}m{r}^{3}{x}^{4
}+9{b}^{4}{l}^{2}{q}^{4}{x}^{4}+12{b}^{4}{l}^{2}{q}^{2}{r}^{2}{x}^
{4}+15{b}^{4}{l}^{2}{r}^{4}{x}^{4}+{b}^{4}{q}^{4}{r}^{2}{x}^{4}-2{
	b}^{4}{q}^{2}{r}^{4}{x}^{4}\nn\\
&+&{b}^{4}{r}^{6}{x}^{4}-12{a}^{5}{b}^{4}l{
	x}^{3}+40{a}^{3}{b}^{4}{l}^{3}{x}^{3}-8{a}^{3}{b}^{4}l{m}^{2}{x}^{
	3}+80{a}^{3}{b}^{4}lmr{x}^{3}-24{a}^{3}{b}^{4}l{q}^{2}{x}^{3}-24
{a}^{3}{b}^{4}l{r}^{2}{x}^{3}\nn\\
&-&28a{b}^{4}{l}^{5}{x}^{3}-8a{b}^{4}{l
}^{3}{m}^{2}{x}^{3}-64a{b}^{4}{l}^{3}mr{x}^{3}+40a{b}^{4}{l}^{3}{q
}^{2}{x}^{3}+8a{b}^{4}{l}^{3}{r}^{2}{x}^{3}-72a{b}^{4}l{m}^{2}{r}^
{2}{x}^{3}\nn\\
&+&56a{b}^{4}lm{q}^{2}r{x}^{3}+48a{b}^{4}lm{r}^{3}{x}^{3}-
12a{b}^{4}l{q}^{4}{x}^{3}-8a{b}^{4}l{q}^{2}{r}^{2}{x}^{3}-12a{b}
^{4}l{r}^{4}{x}^{3}+{a}^{6}{b}^{4}{x}^{2}-28{a}^{4}{b}^{4}{l}^{2}{x}
^{2}\nn\\
&+&8{a}^{4}{b}^{4}{m}^{2}{x}^{2}-12{a}^{4}{b}^{4}mr{x}^{2}+4{a
}^{4}{b}^{4}{q}^{2}{x}^{2}+4{a}^{3}{b}^{3}qr{x}^{4}+37{a}^{2}{b}^{
	4}{l}^{4}{x}^{2}+12{a}^{2}{b}^{4}{l}^{2}{m}^{2}{x}^{2}+84{a}^{2}{b
}^{4}{l}^{2}mr{x}^{2}\nn\\
&-&38{a}^{2}{b}^{4}{l}^{2}{q}^{2}{x}^{2}-18{a}^
{2}{b}^{4}{l}^{2}{r}^{2}{x}^{2}+36{a}^{2}{b}^{4}{m}^{2}{r}^{2}{x}^{2
}-20{a}^{2}{b}^{4}m{q}^{2}r{x}^{2}-12{a}^{2}{b}^{4}m{r}^{3}{x}^{2}
+4{a}^{2}{b}^{4}{q}^{4}{x}^{2}\nn\\
&+&6{a}^{2}{b}^{4}{q}^{2}{r}^{2}{x}^{2
}-3{a}^{2}{b}^{4}{r}^{4}{x}^{2}-4a{b}^{3}{l}^{2}qr{x}^{4}-8a{b}^
{3}mq{r}^{2}{x}^{4}+4a{b}^{3}{q}^{3}r{x}^{4}+4a{b}^{3}q{r}^{3}{x}^
{4}+6{b}^{4}{l}^{6}{x}^{2}\nn\\
&+&24{b}^{4}{l}^{4}mr{x}^{2}-6{b}^{4}{l}
^{4}{q}^{2}{x}^{2}-30{b}^{4}{l}^{4}{r}^{2}{x}^{2}-8{b}^{4}{l}^{2}m
{r}^{3}{x}^{2}+12{b}^{4}{l}^{2}{q}^{2}{r}^{2}{x}^{2}-6{b}^{4}{l}^{
	2}{r}^{4}{x}^{2}+2{b}^{4}{q}^{2}{r}^{4}{x}^{2}\nn\\
&-&2{b}^{4}{r}^{6}{x}^
{2}+6{a}^{5}{b}^{4}lx-28{a}^{3}{b}^{4}{l}^{3}x-8{a}^{3}{b}^{4}l{
	m}^{2}x-24{a}^{3}{b}^{4}lmrx+12{a}^{3}{b}^{4}l{q}^{2}x+12{a}^{3}
{b}^{4}l{r}^{2}x-8{a}^{2}{b}^{3}lmq{x}^{3}\nn\\
&+&16{a}^{2}{b}^{3}lqr{x}^
{3}-10a{b}^{4}{l}^{5}x-24a{b}^{4}{l}^{3}mrx+4a{b}^{4}{l}^{3}{q}^
{2}x+36a{b}^{4}{l}^{3}{r}^{2}x+8a{b}^{4}lm{r}^{3}x-4a{b}^{4}l{q}
^{2}{r}^{2}x\nn\\
&+&6a{b}^{4}l{r}^{4}x-8{b}^{3}{l}^{3}mq{x}^{3}-24{b}^{
	3}lmq{r}^{2}{x}^{3}+8{b}^{3}l{q}^{3}r{x}^{3}+16{b}^{3}lq{r}^{3}{x}
^{3}+9{a}^{4}{b}^{4}{l}^{2}+4{a}^{4}{b}^{4}{m}^{2}+4{a}^{4}{b}^{
	4}mr\nn\\
&+&{a}^{4}{b}^{4}{r}^{2}-2{a}^{4}{b}^{2}{x}^{4}+8{a}^{3}{b}^{3}m
q{x}^{2}-8{a}^{3}{b}^{3}qr{x}^{2}+6{a}^{2}{b}^{4}{l}^{4}+12{a}^{
	2}{b}^{4}{l}^{2}mr+4{a}^{2}{b}^{4}m{r}^{3}+
2{a}^{2}{b}^{2}{l}^{2}{x}^{4}\nn\\
&+&2{a}^{2}{b}^{4}{r}^{4}+4{a}^{2}{b}^{2}mr{x}^{4}-2{a}^{2}{
	b}^{2}{q}^{2}{x}^{4}-2{a}^{2}{b}^{2}{r}^{2}{x}^{4}+16a{b}^{3}{l}^{
	2}mq{x}^{2}-8a{b}^{3}{l}^{2}qr{x}^{2}+24a{b}^{3}mq{r}^{2}{x}^{2}\nn\\
&-&4
a{b}^{3}{q}^{3}r{x}^{2}-8a{b}^{3}q{r}^{3}{x}^{2}+{b}^{4}{l}^{6}+7
{b}^{4}{l}^{4}{r}^{2}+7{b}^{4}{l}^{2}{r}^{4}+{b}^{4}{r}^{6}-8{a}
^{3}{b}^{2}l{x}^{3}-16{a}^{2}{b}^{3}lmqx+8a{b}^{2}{l}^{3}{x}^{3}\nn\\
&+&
16a{b}^{2}lmr{x}^{3}-8a{b}^{2}l{q}^{2}{x}^{3}-8a{b}^{2}l{r}^{2}{
	x}^{3}-16{b}^{3}{l}^{3}qrx+2{a}^{4}{b}^{2}{x}^{2}+8{a}^{3}{b}^{3
}mq+4{a}^{3}{b}^{3}qr-16{a}^{2}{b}^{2}{l}^{2}{x}^{2}\nn\\
&-&8{a}^{2}{b}
^{2}mr{x}^{2}+4{a}^{2}{b}^{2}{q}^{2}{x}^{2}+12a{b}^{3}{l}^{2}qr+4
a{b}^{3}q{r}^{3}+6{b}^{2}{l}^{4}{x}^{2}+16{b}^{2}{l}^{2}mr{x}^{2
}-2{b}^{2}{l}^{2}{q}^{2}{x}^{2}-12{b}^{2}{l}^{2}{r}^{2}{x}^{2}\nn\\
&+&6
{b}^{2}{q}^{2}{r}^{2}{x}^{2}-2{b}^{2}{r}^{4}{x}^{2}+8{a}^{3}{b}^{2
}lx-8a{b}^{2}{l}^{3}x-16a{b}^{2}lmrx-4a{b}^{2}l{q}^{2}x+8a{b}^
{2}l{r}^{2}x+6{a}^{2}{b}^{2}{l}^{2}+4{a}^{2}{b}^{2}mr\nn\\
&+&4{a}^{2}{b
}^{2}{q}^{2}+2{a}^{2}{b}^{2}{r}^{2}-4abqr{x}^{2}+2{b}^{2}{l}^{4}
+4{b}^{2}{l}^{2}{r}^{2}+2{b}^{2}{r}^{4}-8blqrx+{a}^{2}{x}^{2}+4
abqr+2alx+{l}^{2}+{r}^{2}  \}.\nn\\
&&\label{g3}
\end{eqnarray}
The function $Z$ in eq. (\ref{g3}) above is
\begin{eqnarray}
Z&=&{a}^{4}{x}^{4}-{a}^{2}{l}^{2}{
	x}^{4}-2{a}^{2}mr{x}^{4}+{a}^{2}{q}^{2}{x}^{4}+{a}^{2}{r}^{2}{x}^{4}
+4{a}^{3}l{x}^{3}-4a{l}^{3}{x}^{3}-8almr{x}^{3}+4al{q}^{2}{x}^
{3}\nn\\
&+&4al{r}^{2}{x}^{3}-{a}^{4}{x}^{2}+8{a}^{2}{l}^{2}{x}^{2}
+4{a}
^{2}mr{x}^{2}-2{a}^{2}{q}^{2}{x}^{2}-3{l}^{4}{x}^{2}-8{l}^{2}mr{
	x}^{2}+4{l}^{2}{q}^{2}{x}^{2}+6{l}^{2}{r}^{2}{x}^{2}\nn\\
&+&{r}^{4}{x}^{2
}-4{a}^{3}lx+4a{l}^{3}x+8almrx-4al{q}^{2}x-4al{r}^{2}x-3{a
}^{2}{l}^{2}-2{a}^{2}mr+{a}^{2}{q}^{2}-{a}^{2}{r}^{2}-{l}^{4}\nn\\
&-&2{l}
^{2}{r}^{2}-{r}^{4}.
\end{eqnarray}

\end{document}